\documentclass[aps,pre,superscriptaddress,showpacs,nofootinbib] {revtex4}

\usepackage{epsfig}
\usepackage{bm}
\usepackage{graphicx}
\usepackage{amsmath}
\usepackage{amssymb}

\newtheorem{property}{Property}

\begin{document}

\title{Drift of Particles in Self-Similar Systems and its  Liouvillian  Interpretation}
\author{Felipe Barra}
\affiliation{Deptartamento de F\'{\i}sica, Facultad de Ciencias
F\'{\i}sicas y Matem\'aticas, Universidad de Chile, Casilla 487-3,
Santiago Chile}
\author{Thomas Gilbert}
\affiliation{Center for Nonlinear Phenomena and Complex Systems,
  Universit\'e Libre  de Bruxelles, C.~P.~231, Campus Plaine, B-1050
  Brussels, Belgium}
\author{Mauricio Romo}
\affiliation{Deptartamento de F\'{\i}sica, Facultad de Ciencias
F\'{\i}sicas y Matem\'aticas, Universidad de Chile, Casilla 487-3,
Santiago Chile}
\date{\today}
\begin{abstract}
We study the dynamics of classical particles in different classes of
spatially extended self-similar systems, consisting of (i) a self-similar
Lorentz billiard channel, (ii) a self-similar graph, and (iii) a master
equation. In all three systems the particles typically drift at constant
velocity and spread ballistically. These transport properties are analyzed
in terms of the spectral properties of the operator evolving the
probability densities. For systems (i) and (ii), we explain the drift  
from the properties of the Pollicott-Ruelle resonance spectrum and
corresponding eigenvectors
\end{abstract}
\maketitle

\section{Introduction\label{secIntro}}

Billiard systems have long served as paradigm models to study the
foundations of statistical mechanics in connection to ergodic
theory\cite{Arnold-Avez,Sinai70}. 
In the recent years the study of transport 
properties of
ensembles of particles in spatially extended systems, like diffusion  in the
Lorentz gas or multi-baker map \cite{Gasp,tasaki} and heat  conduction in
similar systems \cite{alonso1,alonso2,Leyvraz},  has proven 
to be very fruitful in establishing connections between irreversible
phenomena at the macroscopic scales and the chaotic properties of the
reversible classical dynamics at the microscopic scales
\cite{GaspB,Dorfman}.

In this respect an essential tool is the Liouvillian formulation of the
dynamics.  In this formulation, instead of considering the behavior of
individual trajectories, we consider the evolution of a density $\rho_0(X)$
of initial conditions.  This density evolves in phase space  according to
the Liouville equation $\partial_t \rho_t(X) +\hat{L}\rho_t(X)=0$, where
$\rho_t(X)$ represents the density at time $t$ in phase space of points $X$
and the operator $\hat{L}=\{H,\cdot\}$ with $H$ the Hamiltonian of the
system and $\{\cdot,\cdot\}$ the Poisson bracket is called the Liouvillian
operator. This equation is integrated using the initial condition
$\rho_0(X)$ \cite{Goldstein}.  We write its solution in the form
$\rho_t(X)=(\hat{P}^t \rho_0)(X)$ where we introduced the evolution
operator $\hat{P}^t $ known as the Perron-Frobenius operator. When we are
interested in the future time evolution of the system we may analyze this
operator in terms of the Pollicott-Ruelle resonance spectrum $\{s_j\}$ of
the chaotic systems \cite{Ruelle}, which determines the decay rates of the
system. That is, for long times, the density can be decomposed on modes
which decay exponentially in time $\rho(X,t)\sim \sum_j e^{s_j t}c_j
\psi_j(X)$ with the $c_j$ determined by the initial condition. Within this
theoretical framework, macroscopic properties have been related to
microscopic quantities. In particular it has been applied to
billiard systems which are spatially periodic and whose extension is
infinite in one or two directions \cite{GaspB}. For these systems,
analytical results 
can be obtained using the Bloch theorem. In fact, expanding the functions
like densities $\rho$ and eigenstates of the Liouville operator $\psi_j$ in
Fourier series it is  possible to analyze the problem in the finite domain
of a unit cell instead of the infinite domain of the extended  billiard,
and compute 
quantities like eigenvalues as function of the wavenumber.

In studying the application of this formalism to situations of physical
interest, an important problem concerns the characterization of classes of
systems other than spatially periodic ones, where one can successfully
apply the spectral theory of the evolution operator.

Here we study a class of extended billiard models with a self-similar
structure and show how the techniques described above can be transposed to
such systems. By self-similar billiard, we mean a billiard made up of a
collection of cells with a one-dimensional lattice structure, where the
cell sizes increase exponentially with their indices. Because of
this property, particles will move in a preferred direction and therefore
have a mean drift. That is, contrary to the periodic case, a 
density of particles drifts with constant velocity and does not diffuse. It
is our goal to provide a theoretical understanding of these properties,
based upon the spectral analysis of the evolution (Perron-Frobenius)
operator of the system.

By understanding we mean that, although we do not obtain explicit
solutions for the eigenvalues and eigenvectors of the evolution  operator,
we show that they verify  two properties which are essential to produce
this drift. Thus, we establish the connection  between a transport
property of macroscopic nature and the evolution operator acting on
phase-space trajectories, for a new class of  spatially extended
systems.
 
A comparison between different levels of description is achieved by
considering successively a fully deterministic self-similar billiard, then
introducing a mesoscopic model with stochastic collision rules, and finally
a marcoscopic model in the form of a master equation.

The article is organized as follows. Section \ref{secHB} describes the
self-similar billiard and the evolution of a particle in it. 
We identify two properties which characterize the spectrum
of the evolution operator and discuss their consequences on the evolution
of statistical ensembles. 
In order to help understand these features, we introduce, in
Sec. \ref{secHG}, a class of self-similar graphs.
We show this system verifies two properties similar to those of the
billiard. 
In Sec. \ref{SecMeq}, a phenomenological  approach is given, based on a
Master equation, 
for which we obtain exact expressions for the drift velocity and the mean
square displacement. This provides 
theoretical predictions for the billiard and the graph models which 
are compared to numerical computations.
Conclusions are drawn in Sec. \ref{conclu}.

\section{Self-similar billiards\label{secHB}}

We will consider self-similar billiard chains such as shown in
Fig.~\ref{Fig2}, which consist of an infinite
collection of two-dimensional cells, shown in Fig.~\ref{Fig1}, glued
together along a horizontal 
axis. Each cell contains convex scatterers and is open so as to allow
particles to flow from one cell to the next. The shapes of the cells are
identical, but their sizes are taken to grow exponentially with their
indices. The overall geometry is such that upon combined shifting and
rescaling the whole billiard is unchanged.

\subsection{Definition of the Model}

\begin{figure}[htb]
\centering
\hspace{-1cm}
\includegraphics[height=7cm]{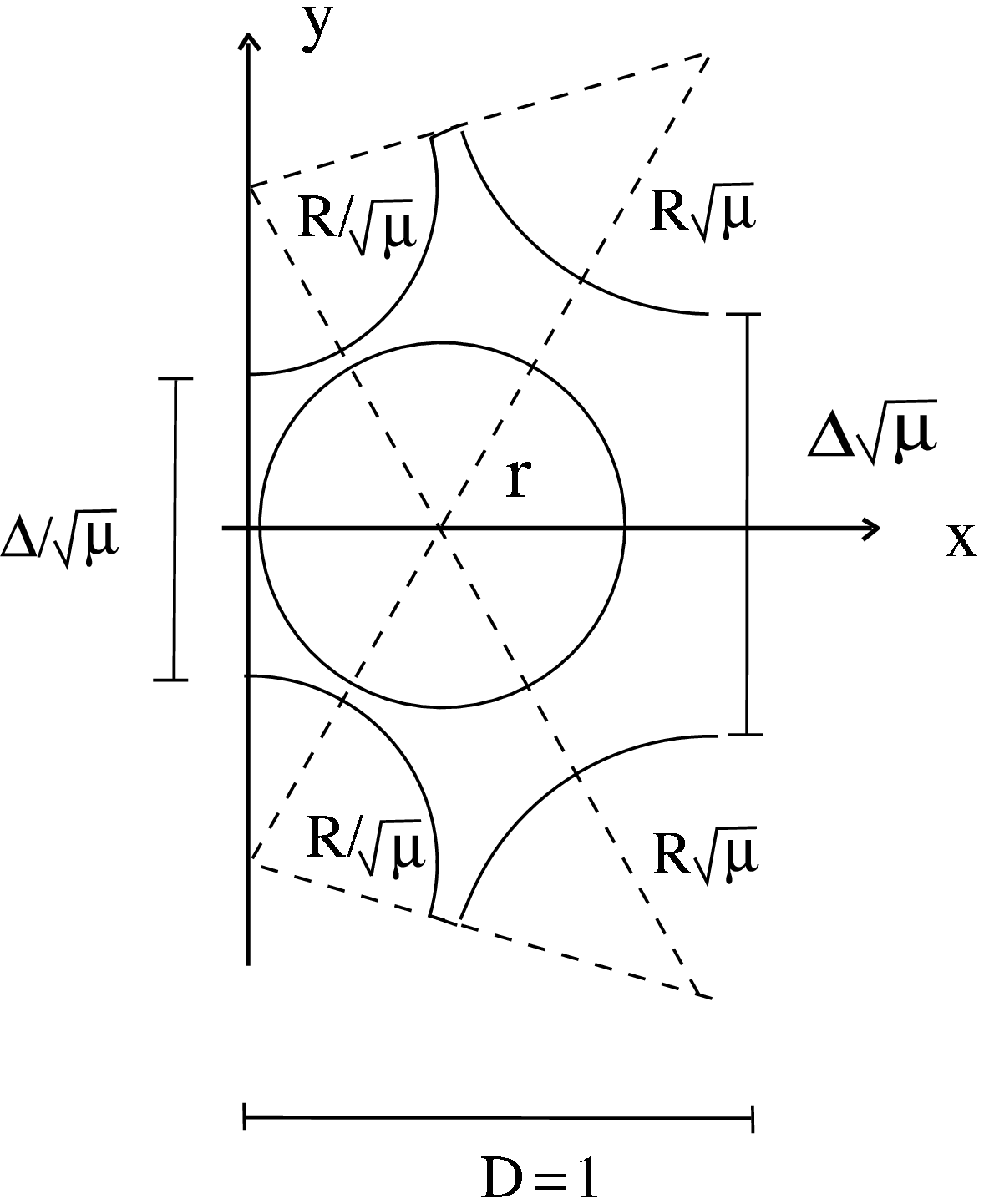}
\hspace{1cm}
\includegraphics[height=7cm]{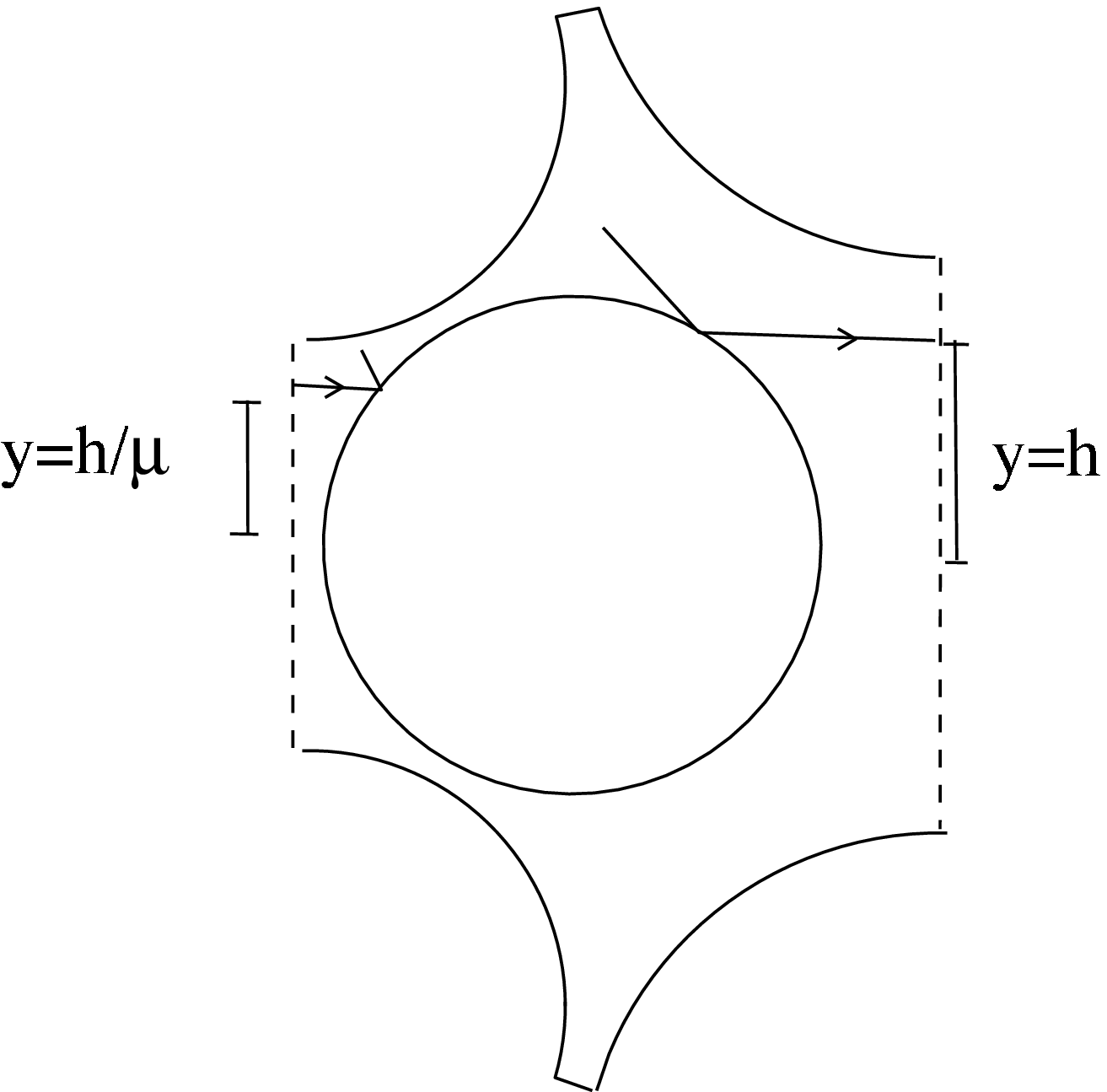}
\caption{Left Panel : The reference cell geometry. A detailed description is presented in appendix
\ref{Appgeomcell}. Here it is sufficient to note that $\Delta=\sqrt{3}D-2R$ and that with this choice
the usual Lorentz channel is retrieved for $\mu=1$. Right Panel : The arrows on the right pannel
  represent schematically the  matching condition~: A trajectory  leaving a
  cell from the right is reinjected to the left with a change of  vertical coordinate
  $y=h\to y=\frac{h}{\mu}$ and the velocity changed according to
  $v  \to \frac{v}{\mu}$.}
\label{Fig1}
\end{figure}

We consider a self-similar billiard chain based on the
Lorentz channel \cite{GaspB}. 
The reference cell is represented in Fig.~\ref{Fig1}.
It is a region defined by the
exterior of five disks, four of which are half disks, located at the
corners of the cell and shared with the neighboring cells, and one located
at the center of the cell. The dissymmetry  between 
the left- and the right-hand sides depends on the scaling  parameter
$\mu$ ($\mu=1$ is the symmetric case). Given the value of $\mu$, there are
three other parameters, namely $D$, $R$ and $r$, which, as shown in
Fig.~\ref{Fig1}, determine respectively the horizontal width of the
reference cell, the radii of the external disks and the radius of the
center disk. Of these three pamarameters, only the ratios $R/D$ and $r/D$
are actually relevant. Notice the mirror symmetry of the unit cell about its 
center under the transformation $\mu\to1/\mu$.
Appendix \ref{Appgeomcell} details the restrictions
imposed on the values of the parameters, chosen so that the self-similar
billiard shares the hyperbolicity of the Lorentz channel.

The two vertical segments of lengths $\Delta/\sqrt{\mu}$ and
$\Delta\sqrt{\mu}$, with
$\Delta\equiv\sqrt{3}D-2R$, at the left and right boundaries 
will be referred to as 
the windows of the cell, because a particle that goes across them moves
from one cell to one of its neighbors, as will be detailed  below.

It will be convenient to take $D=1$ and rescale $R$ and $r$
accordingly. Thus our billiard is characterized by $\mu$, $R/D$ and $r/D$.

\begin{figure}
\includegraphics[width=.5\textwidth]{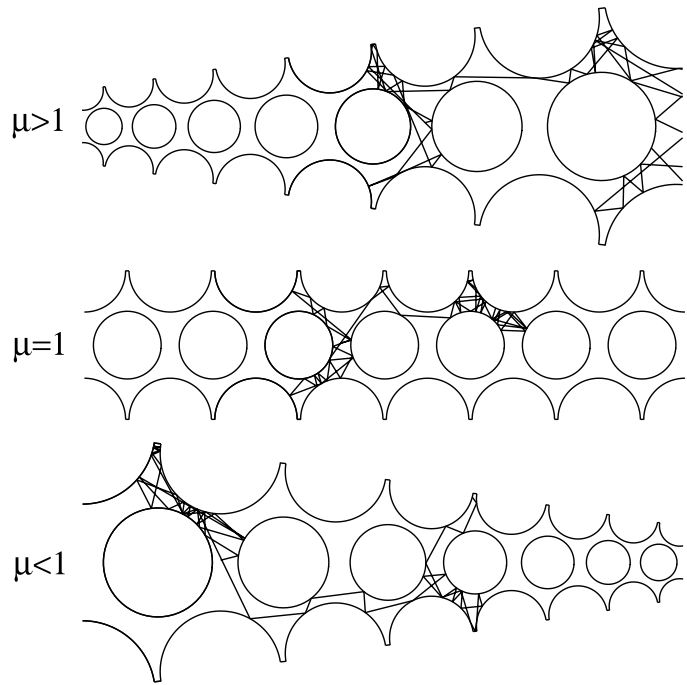}
\caption{The self-similar billiard and a trajectory for $\mu>1$,  $1/ \mu<1$   and
  $\mu=1$ respectively. Notice the symmetry of the billiard  under $\mu \to
  1/\mu$ and $x\to -x$.}
\label{Fig2}
\end{figure}

The whole chain is constructed by adding a cell to the right of
the reference cell, identical in shape but with all its lengths  multiplied
by $\mu$ and another one to the left with  all the lengths divided by
$\mu$. We repeat this construction in such a way that in the $i$th cell to
the right all the lengths are multiplied by $\mu^i$ and, equivalently,  by
$\mu^{-i}$ to the left. The resulting billiard chain depicted in
Fig.~\ref{Fig2}, is so constructed that the mirror symmetry with 
respect to the transformation $\mu\to1/\mu$ remains.

Now we consider a particle moving inside the billiard with velocity
$\vec{v}$. Figure \ref{Fig2} shows such a trajectory.
As the particle moves from one cell to a neighboring one, the length scales
change by a factor $\mu$, so that the characteristic time between
collisions with the walls changes accordingly (the speed stays constant).
Equivalently we can rescale the velocity by $\mu$ and keep the  length  scales
unchanged. That is, in going from one billiard cell to the next, say
from left to right, the following transformations are equivalent~:
\begin{equation}
\left\{
\begin{array}{c}
\vec{v}\to \vec{v}\\
l\to \mu l
\end{array}
\right\}
\Leftrightarrow
\left\{
\begin{array}{c}
\vec{v}\to \frac{\vec{v}}{\mu}\\
l\to l
\end{array}
\right\}.
\label{mapping}
\end{equation}
In both transformations the time between collisions with the walls is
shorter ($\mu<1$) or longer ($\mu>1$) by a factor  $\mu$.

Therefore we can analyze the dynamics on the self-similar billiard  in  terms
of the dynamics in a periodic billiard if, instead of rescaling the  size of
the cell, we rescale the velocity. This way, the dynamics on the
infinite self-similar billiard  chain can be reduced to the dynamics  on a
single cell, provided the matching conditions given in Table \ref{table1}
are imposed.
\begin{table}[h]
\begin{tabular}{|llc||c|c|}
\hline
&&&Exit to the right&Exit to the left\\
\hline
position & $\vec{r}$&:& 
$(D,h)\quad\to\quad(0,h/\mu)$ &
$(0,h')\quad\to\quad(D,h'\mu)$ \\
velocity & $ \vec{v}$ &:& 
$(v_x,v_y)\quad \to \quad(v_x/\mu,v_y/\mu)$ 
& $(v_x,v_y)\quad \to \quad(v_x \mu,v_y\mu)$\\
\hline
\end{tabular}
\caption{\label{table1}Matching conditions for trajectories
crossing from one side of the cell to the other, as depicted on the
right pannel of Fig.~\ref{Fig1}. The notations $h$ and $h'$ are detailed in
the text.
}
\end{table}

In Table \ref{table1}, $h$ is the $y$ coordinate of the trajectory  escaping
through the right  window, $-\sqrt{\mu}\Delta/2<h<\sqrt{\mu}\Delta/2$  and
$h'$, that of the  trajectory escaping through the left window,
$-\Delta/(2\sqrt{\mu})<h'<\Delta/(2\sqrt{\mu})$. For the collisions   with
the walls the dynamics is determined by the Birkhoff map  \cite{GaspB}
$(s,v_\mathrm{t})_n\to(s,v_\mathrm{t})_{n+1}$ of our modified Lorentz
channel. Here the variable $s$ represents the arc-length along the unit
cell boundary and  $v_\mathrm{t}$ the projection of the  
normalized velocity ($|v|=1$) to the vector tangent to the boundary. This map,
together with  Table \ref{table1}, provides the map for the evolution of a
particle in the  self-similar billiard, using only one
cell. The matching conditions given in table \ref{table1} are analogous to
periodic  boundary conditions for a periodic billiard without the self-similar
structure of our billiard. 

The construction of the self-similar billiard we have
presented is  general and can be used to construct other self-similar
billiards by  changing the choice of the unit cell. The crucial point is that
the cells are scaled uniformly by the factor $\mu$ at every step of the
hierarchy. 

To our knowledge, we are the first to consider a billiard with such a geometry.

\subsection{Poincar\'e Map}

As we have shown, the evolution on a self-similar billiard can be
considered  on a single cell, provided we change the  speed of
the particle at every time it crosses the windows. The  mapping from one
point of the boundary to the next can be described by the variables $s $ and
$v_\mathrm{t}$. Let $\xi$ denote this pair of variables. The
map,
\begin{equation}
\xi_{n+1}=\phi(\xi_n),
\label{poincare}
\end{equation}
which determines the sequence of points visited on the boundary and the
corresponding projection of the normalized velocity to the tangent vector,
is called the "Poincar\'e 
map". The area where $\xi$ lives defines the Poincare surface of section
$\mathcal{P}$. 

This Poincar\'e map misses the information on the speed, which we 
can restore as follows. To consider the change of speed we need to keep
track of  the cell where the particle is located after the $n$
iterations. Let us introduce  a new 
variable $I_n$, which takes integer values and labels the cell where the
particle is at the $n$th iteration of the map. We defined the jump function
$a(\xi)$, such that
\begin{equation}
I_{n+1}=I_n+a(\xi_n),
\label{jump}
\end{equation}
where $a(\xi_n)=1$ if  $\phi(\xi_n)$ has the spatial coordinate
$s_{n+1}$ on the right window, and $a(\xi_n)=-1$ if $\phi(\xi_n)$ has the
spatial coordinate $s_{n+1}$ on the left window. Otherwise $a(\xi_n)=0$.

Using the variable $I$ we can determine the actual speed from
Eq.~(\ref{mapping}). Equivalently, we can say that the time it takes a
trajectory at a  (phase space) point $\xi$ on the boundary to intersect
again with the boundary of the billiard depends on $I$ as
\begin{equation}
T(\xi,I)=\frac{L(\xi)}{v}\mu^I
\label{Time}
\end{equation}
where $L(\xi)$ is the length of the trajectory between intersections at
$\xi$ and $\phi(\xi)$ of the trajectory with the boundary of the unit  cell
and  $v$ is the speed of the particle.

Now we have a complete description of the dynamics in the self-similar
billiard. Every point of phase space is characterized by $X=[\xi, \tau,I]$
with  $0<\tau<T(\xi,I)$ a new variable that restores the position
between collisions. The complete flow $\Phi^t(X)$ can be specified using
these coordinates. This decomposition of the flow in terms  of  a
Poincar\'e map  
and the variable $\tau$ along the trajectory issued from the Poincar\'e section
is called a suspension flow. It can be implemented in billiards and other
type of systems \cite{Sinai}.

\subsection{Evolution of Statistical Ensembles}

Consider now an arbitrary distribution of initial conditions. The evolution
of this statistical ensemble is determined by the Perron-Frobenius operator
\begin{equation}
(\widehat{P}^t\rho)(X)=\rho(\Phi^{-t}X) \ ,
\label{PFO}
\end{equation}
which requires knowledge of the backward dynamics given by
\begin{equation}
\Phi^{-t}(\xi,\tau,I) =
\left\{
\begin{array}{l}
(\xi,\tau-t,I), \qquad {\rm if}\quad 0 \leq t < \tau,\\
(\phi^{-1}\xi, \tau - t + T(\phi^{-1}\xi, I - a(\phi^{-1}\xi)), I -
a(\phi^{-1}\xi)), \\
\qquad {\rm if} 
\quad \tau\leq t < \tau + T(\phi^{-1}\xi, I - a(\phi^{-1}\xi)),\\
\vdots
\end{array}
\right.
\label{phit}
\end{equation}
At $t=\tau$ we cross the section $\mathcal{P}$ and
we have to identify
$[\xi,0,I]=[\phi^{-1}\xi, T(\phi^{-1}\xi, I-a(\phi^{-1}\xi)),
I-a(\phi^{-1}\xi)]$.
The interpretation is the following~: for a particle in the cell $I$,
its position and velocity are completely specified by $\xi$, which provides
both the direction of the velocity and the last point of intersection  with
the perimeter of the cell. Moving a distance $\tau$ along the line going
from this  point of intersection in the direction of the velocity (both
specified by $\xi$) we retrieve the exact position of the particle. 

Now running the time backwards at $t=\tau$ we are just at the point  of
intersection with the billiard. This point is also the end of the segment
issued from $\phi^{-1}(\xi)$, the previous intersection with the perimeter
and at $\tau = T(\phi^{-1}\xi, I - a(\phi^{-1}\xi))$ which is the upper
limit for the possible values of $\tau$ that start at $\phi^{-1}(\xi)$,
with the  corresponding direction. 
This intersection is not necessarily in the same cell $I$. The
dependence on $I - a(\phi^{-1}\xi)$ indicate that the point $\phi^{-1}(\xi)$
is in the cell $I - a(\phi^{-1}\xi)$. This identification is made at every
point of intersection. 

In general, we can write Eq.~(\ref{phit}) under the compact form,
\begin{eqnarray}
\Phi^{-t}[\xi,\tau,I]&=&\left[\phi^{-n}
  \xi,\tau-t+\sum_{i=1}^nT\left(\phi^{-i}\xi,I-\sum_{j=1}^{i}
    a(\phi^{-j}\xi)\right),I-\sum_{i=1}^{n}a(\phi^{-i}\xi)\right],
\label{Eq.8}
\\
&&\mathrm{if} \quad  0< \tau-t + \sum_{i=1}^nT
\left(\phi^{-i}\xi, I - \sum_{j=1}^{i} a(\phi^{-j}\xi)\right) < T
\left(\phi^{-n}\xi, I - \sum_{j=1}^{n}a(\phi^{-j}\xi)\right)
\nonumber
\end{eqnarray}

This construction introduces a small generalization of the treatment
of periodic billiards described in \cite{Gas96}. More details can be
found in that article. 

This suspension flow formalism of the billiard dynamics  has two main
advantages~: (i) it provides a method to  simulate 
the dynamics of particles in extended billiards and, (ii) it can be  used
(at least formally) to determine the spectrum of the evolution  operator as
shown in \cite{Pollicott,Gas96}. The generalization is direct, the only
difference being the dependence of $T$ on $I$, the cell index. 

Next we present the results relevant to
this generalization and deduce two properties of the spectrum  of the
Perron-Frobenius operator that follow from the self-similar structure.

\subsection{Two Properties of the Spectrum of the self-similar Billiard}
\label{b-bill}

Following \cite{Gas96}, by a Laplace transform of the Perron-Frobenius
operator Eq.~(\ref{PFO}) and using Eq.~(\ref{Eq.8}), we obtain the following equation for the
Pollicott-Ruelle resonance spectrum $\{s\}$ and the associated  eigenstates
$\{b_s\}$,
\begin{equation}
R_{s}b_{s}[\xi,I]=b_s[\xi,I],
\end{equation}
with $R_{s}$ defined by
\begin{equation}
R_{s}f_{s}[\xi,I] = \exp[-sT(\phi^{-1}\xi, I - a(\phi^{-1}\xi))]
f_{s}(\phi^{-1}\xi, I - a(\phi^{-1}\xi)).
\end{equation}
In this expression, the operations must be understood in the following
order~:  First take $\xi \to \phi^{-1}\xi$ everywhere and then $I \to
I-a(\phi^{-1}\xi)$.
The operator $R_{s}$ can be considered as a reduced  Perron-Frobenius
operator which evolves densities between successive crossings of the
surface of section. A formal eigenstate can be obtained by successive
applications of this  operator to the identity. That is, 
\begin{eqnarray}
b_s[\xi,I] &\equiv&  \lim_{n\to\infty} R_{s}^{n},\nonumber\\
&=& \prod_{j=1}^{\infty} \exp\left[-sT\left(\phi^{- j}\xi, I - \sum_{i=1}^{j}
  a(\phi^{-i}\xi)\right)\right],\nonumber\\
&=& \prod_{j=1}^{\infty} \exp\left[- \frac{s}{v} L(\phi^{-j}\xi)
  \mu^{I - \sum_{i=1}^{j}a(\phi^{-i}\xi)}\right],
\label{formalb}
\end{eqnarray}
which satisfies
\begin{equation}
R_{s}b_s=b_s.
\end{equation}
Here we used Eq.~(\ref{Time}) to obtain the last equality in
Eq.~(\ref{formalb}). Similarly, an eigenstate $\tilde{b}_s$ of the  adjoint
operator can be  obtained.

Now, we assume that a value of $s$, {\em i.~e.} a Pollicott-Ruelle
resonance, is known and thus the formal expression of $b_s$,
Eq.~(\ref{formalb}), defines the 
corresponding eigenstate. We want to show that, given the pair $s$ and
$b_s[\xi,I]$ (normalized) that satisfies
\begin{equation}
(R_s b_s)[\xi,I]=b_s[\xi,I],
\label{uno}
\end{equation}
there corresponds a pair $s\mu$,  $b_{s\mu}$ such that
\begin{equation}
(R_{s\mu} b_{s\mu})[\xi,I]=b_{s\mu}[\xi,I],
\label{star}
\end{equation}
{\em i.~e.} $s\mu$ is also a resonance. 

In order to show this, we make
the following ansatz~: $b_{s\mu}[\xi,I]=b_s[\xi,I+1]$.
First notice  $b_{s \mu}$
exists and it is normalized by construction. We need to check that
Eq.~(\ref{star}) is satisfied~: 
\begin{eqnarray}
(R_{s\mu} b_{s\mu})[\xi,I] &=& \exp[-s\mu  T(\phi^{-1}\xi, I -
  a(\phi^{-1}\xi))]  b_{s\mu}[\phi^{-1}\xi, I - a(\phi^{-1}\xi)],  
\nonumber\\
&=& \exp[-s\mu T(\phi^{-1}\xi, I - a(\phi^{-1}\xi))]  b_{s}[\phi^{-1} \xi, I +
1 - a(\phi^{-1}\xi)], \nonumber\\
&=& \exp[-sT(\phi^{-1}\xi, I + 1 - a(\phi^{-1}\xi))]  b_{s}[\phi^{-1} \xi,I + 1
- a(\phi^{-1}\xi)], \nonumber\\
&=& (R_{s} b_{s})[\xi, I + 1],\nonumber\\
&=& b_{s}[\xi, I + 1],\nonumber\\
&=& b_{s\mu}[\xi, I] \ .
\label{dem}
\end{eqnarray}
The first equality follows from the definition of $R_s$, the second  from  the
ansatz for $b_{s\mu}$, the third from the scaling property of $T$, the
fourth using again the definition of $R$, the fifth from Eq.~(\ref {uno})
and the sixth again from the definition of $b_{s\mu}$.
Hence $s\mu$ is a resonance and $b_{s\mu}[\xi,I]=b_s[\xi,I+1]$ is the
corresponding eigenstate.

 In the appendix \ref{appendix} we give a proof without this assumption. 

These two properties are essential to what follows~:
\begin{property}[resonances]
\label{propA}
If $s$ is a Pollicott-Ruelle resonance of a
    self-similar billiard, so is $s\mu$.
\end{property}
\begin{property}[eigenstates]
\label{propB}
If the eigenstate associated to $s$ is
    $b_s(\xi,I)$, then the eigenstate $b_{s\mu}(\xi,I)$ associated to
    $s\mu$, is equal to $b_s(\xi,I+1)$.
\end{property}

We end this section with a few remarks. These properties follow
essentially from the exponential dependence of $T(\xi,I)$ on $\mu$ and
thus do not require any assumption on the geometry  of the unit cell apart
from those needed to have a spectral decomposition of the  evolution
operator \cite{Gas96}. Property \ref{propA} says that the mode $b_{s\mu}$
has a  lifetime which is a  factor $1/\mu$ shorter if $\mu>1$ (resp. longer
if $\mu<1$) than the lifetime of the mode $b_{\mu}$, and  property \ref{propB} 
says that,  the mode $b_{s\mu}$  is equal to the mode $b_{\mu}$, but shifted to
the left by one  cell of the  self-similar chain.

\subsection{Transport Properties of self-similar Billiard: Numerical
  Results \label{sec_billnr}}

We claim the two properties derived in the previous section govern the
macroscopic behavior of a density of particles in the  
system. Indeed it will be shown in Sec.~\ref{secDrift} that these
properties induce a constant drift of the particles towards the direction
of growing cells. 
More information on the spectrum, such as shape of the eigenstates and
values of the resonances $s$ are actually needed to compute an explicit
formula for the velocity, but with these two properties we can already
understand the main behavior, {\em i.~e.} the drift. Another important
aspect of the macroscopic evolution is the spreading of particle densities
around the drift. 

In order to illustrate this macroscopic behavior, let us consider an
ensemble of initial conditions 
located in a given cell $I_0$, whose velocities are distributed at random
angles, but with the same magnitude $v_0$. We follow the evolution of this
density and compute the average position along the horizontal axis,
$\langle X \rangle$, of this ensemble as a function of time.
In these simulations we fix two parameters, namely $r/D=0.395$ and 
$R/D=0.480$ an we vary $\mu$ in the interval $0.653<\mu<1.532$ 
determined by the conditions of appendix \ref{Appgeomcell}.

\begin{figure}[ht]
\centering
\includegraphics[width=.4\textwidth]{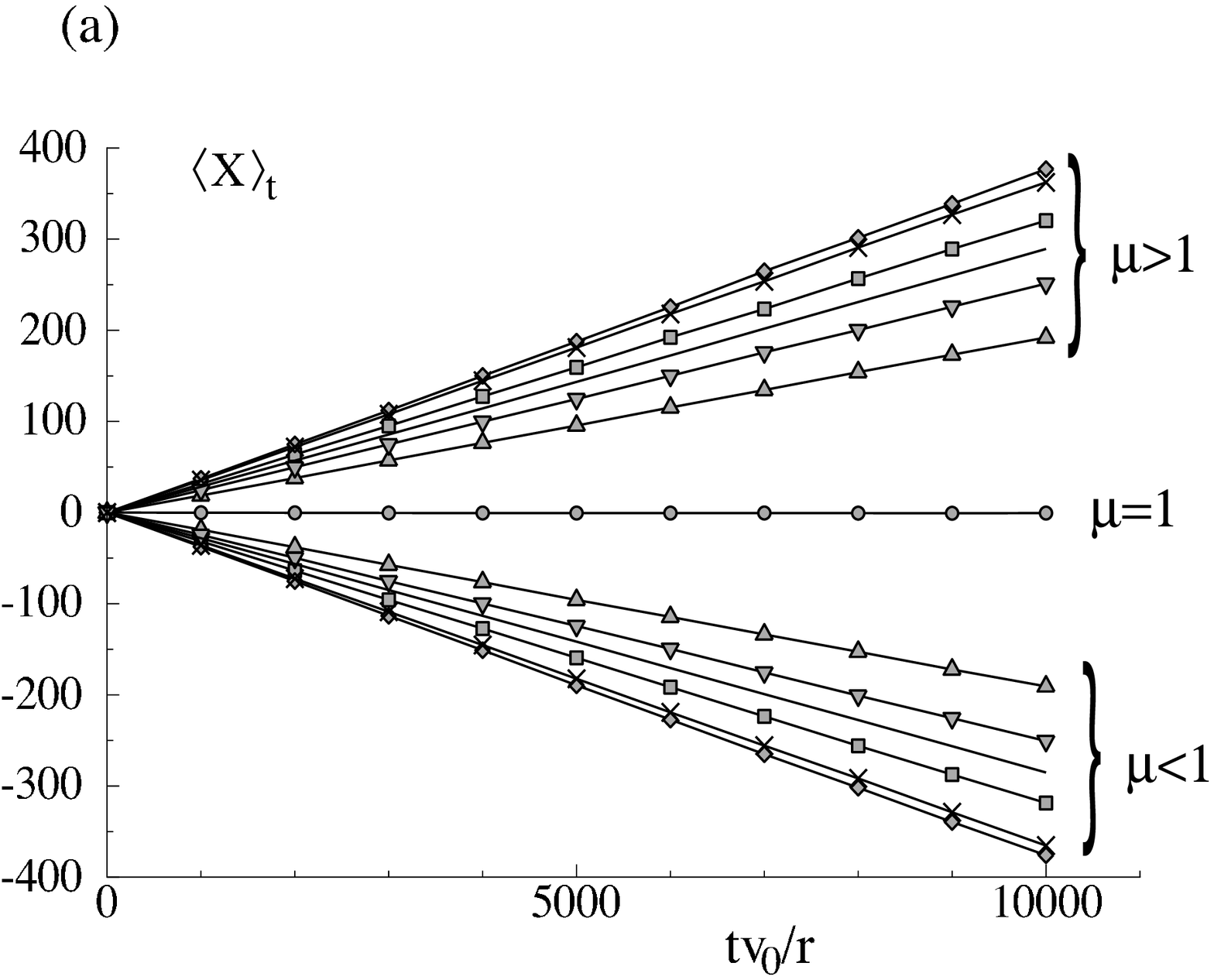}
\includegraphics[width=.4\textwidth]{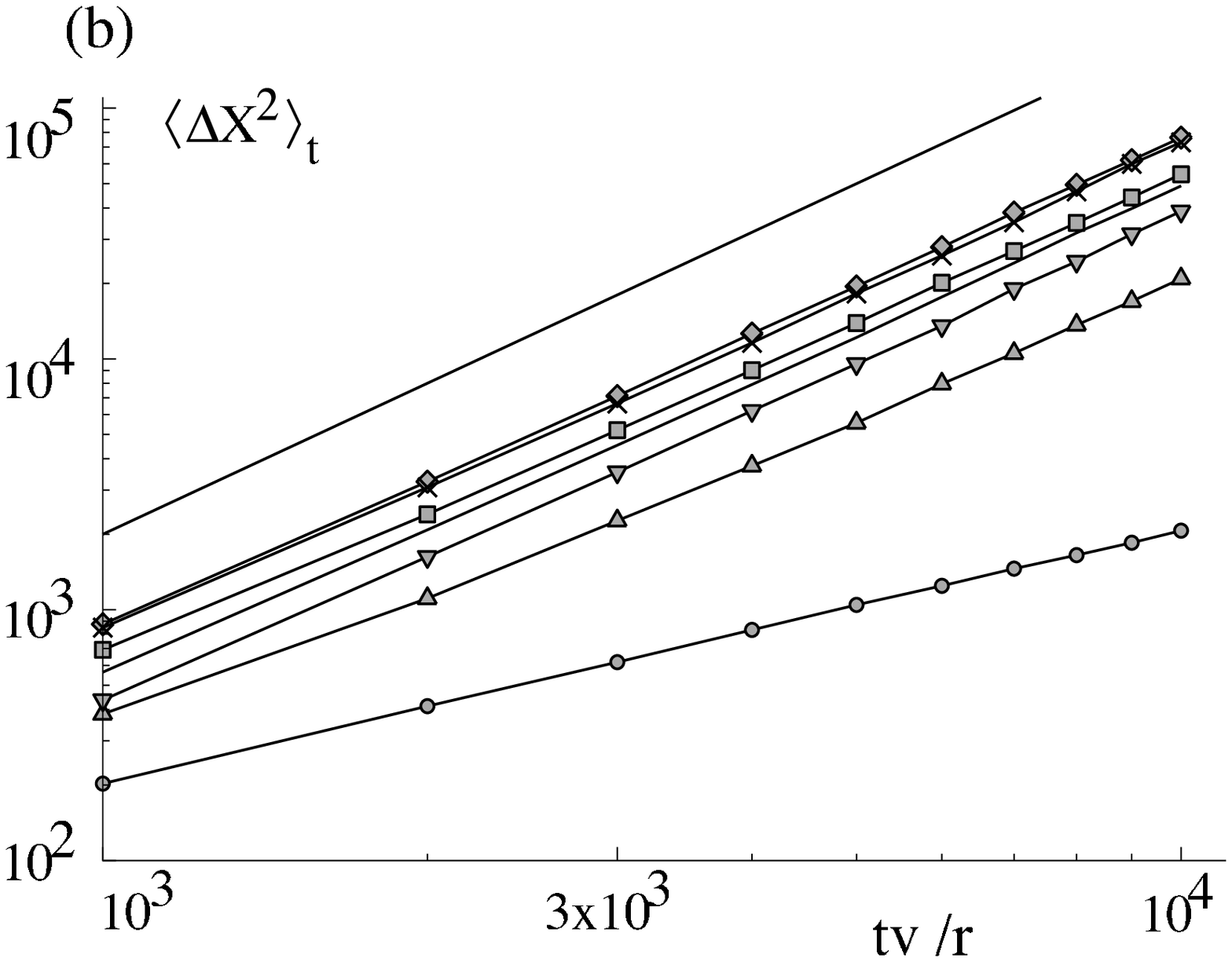}
\caption{(a) Average horizontal position $\langle X\rangle_t$  in the
  billiard, as a function of time for different values of $\mu$. We  see
  that for each value of $\mu$ there is a well defined slope (average
  velocity) and as $\mu$ increases so does the slope. (b) Plot of the mean
  square displacement $\langle \Delta X^2\rangle_t$,  for the same values
  of $\mu$. We observe a clear quadratic behavior $\sim t^2$ (upper solid
  line) for all values of  $\mu$, except for $\mu=1$  (lower curve) where
  the mean square displacement behaves linearly with time. For  panel (a) 
  the data corresponds to  $\mu= 1/1.53$ ($\triangledown$), 1/1.5 ($-$),
  1/1.4 ($\times$), 1/1.3 ($\diamond$), 1/1.2 ($\square$), 1/1.1
  ($\bigtriangleup$), 1 ($\circ$), 1.1 ($\bigtriangleup$), 1.2 ($\square$), 1.3
  ($\diamond$), 1.4 ($\times$), 1.5 ($-$), 1.53 ($\triangledown$). Note
  that  these values are chosen such that we have pairs $\mu$ and
  $1/\mu$. The symmetry $v\to-v$ when $\mu \to 1/\mu$ is seen in  figure
  (a). In  figure (b) the data for $\mu$ and $1/\mu$ almost superpose, thus
  we have plotted  only values for $\mu\geq1$. We recall the parameter
  values are $R/D = 0.48$ and $r/D = 0.395$.}
\label{Fig-vel-billiard}
\end{figure}

For $\mu=1$ the chain is periodic and we know that the mean value stays
constant (no drift) and the spreading of the density is that of a diffusive
process. Thus  the density for long times is distributed according to a
Gaussian, with a variance growing as $\sqrt{t}$. This is confirmed by the
numerical  result shown in Fig.~\ref{Fig-vel-billiard}, central curve on
panel (a) and bottom curve on panel (b).

\begin{figure}[ht]
\centering
\includegraphics[width=.4\textwidth]{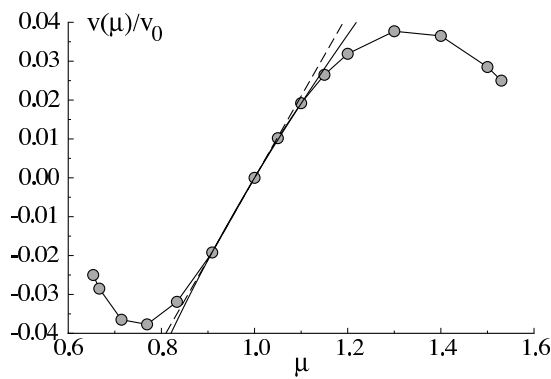}
\caption{Dots: Plot of the velocity $v$ as a function of $\mu$.  The solid line
  is the function $c(\mu-1/\mu)(\sqrt{\mu}+1/\sqrt{\mu})$, with $c=0.05$  a
  constant we use to fit the data close to $\mu=1$ (the reason to  plot
  this function will become clear in Sec.~\ref{SecMeq}). The dashed line is
  a straight line with slope 0.21. Note that for values of $\mu$ away from
  $\mu=1$ the dots are not well approximated by the line.} 
\label{Fig-vel-mu-bill}
\end{figure}

For $\mu \neq 1$ we observe that the packet moves
towards the region of growing size cells ({\em i.~e.} to the left for
$\mu<1$ and 
right for $\mu>1$). The velocity of this motion is obtained by computing
the average position of the particles at different times, which is plotted
in Fig.~\ref{Fig-vel-billiard}. Clearly there is a well defined constant
speed for each value of $\mu$. This speed is plotted as a  function of
$\mu$ in Fig.~\ref{Fig-vel-mu-bill}. The 
data is plotted together with the speed, Eq.~(\ref{vel}), obtained   from
a macroscopic model based on a master equation to be described in
Sec. \ref{SecMeq} and a linear approximation for small values of $\mu-1$. 

We also computed the mean square displacement $\Delta X^2=\langle
X^2\rangle-\langle X\rangle^2$ as a function of time. The numerical
observation, Fig.~\ref{Fig-vel-billiard}b is that $\Delta  X^2\sim t^2$,
{\em i.~e} absence of diffusion, as
expected from the macroscopic description of the  system in
Sec. \ref{SecMeq}.

\section{A self-similar Graph\label{secHG}}

In order to simplify matters a bit, we introduce a stochastic system  in the
form of a graph, with properties similar to the self-similar  billiard
described in the previous section. In a
sense, this self-similar graph is an intermediate description
between the billiard and a macroscopic description based on a Master
equation, to be discussed in Sec.~\ref{SecMeq}.

Graphs are geometrical objects made of vertices connected by bonds of  given
lengths, on which particles are allowed to move according to specific
rules. The particles move freely on the bonds and get randomly  scattered at
the vertices to any of the connected bonds, according to prescribed
transition probabilities. The classical dynamics of graphs share many
properties of  one-dimensional chaotic maps. in fact every graph can be
mapped to an expanding one-dimensional map \cite{ClasGraph}.

The model we consider is self-similar in the sense that the
lengths of the bonds grow exponentially with their indices. Here the chaotic
dynamics of the billiard is replaced by a one-dimensional free flight and a
probabilistic law that determines the transition probabilities between cells.

\subsection{Definition of the Model}

The graph is defined as follows. It consists of
a linear chain of bonds, connected to one another from end to end.
Transition probabilities to the left, $T_L$, and to the right, $T_R$, are  taken to
scale according to $T_R/T_L=\mu$, with the parameter $\mu$ playing a role
similar to the billiard's.
This choice is motivated by the dynamics of the billiard.
Indeed, provided we can assume the escape probabilities are small, we
expect trajectories to distribute uniformly within the cells before
exiting, so that the escape probabilities to the left and right are within
a  ratio $\mu$ of one another. 
Thus let $T_L=p/\sqrt{\mu}$ the transition probability to the
left, and  $T_R=p\sqrt{\mu}$ to the right. Like in the billiard, we  have a
reference cell (with index zero) with a bond whose length we take to be
$L_0$. The first cell to the right has length  $L_1=L_0\mu$ and the $i $th
cell to the right has length $L_i=L_0\mu^i$. Likewise the first cell  to the
left has length $L_{-1}=L_0\mu^{-1}$ and the  $i$th cell to the left
$L_{-i}=L_0\mu^{-i}$. The transition probabilities for all the cells are
the same as for the zeroth cell and we take for left vertex  transition
$T_L$ and reflection $R_L=1-T_L$, and for the right vertex $T_R$ and
reflection $R_R=1-T_R$.

\begin{figure}[ht]
\centering
\includegraphics[width=.4\textwidth]{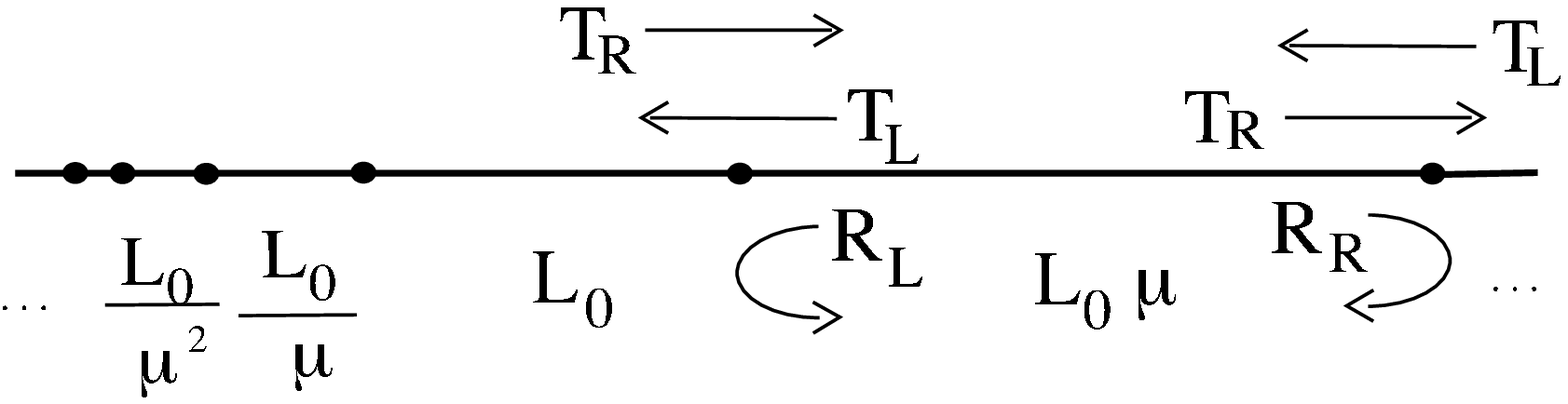}
\caption{Schematic representation of a self-similar graph with $\mu=2$.}
\end{figure}

\subsection{Two Properties of the Spectrum of the self-similar  Graph}
\label{secSG}

We identify two properties of the spectrum of the Perron-Frobenius  operator
of the self-similar graph, analogous to Properties \ref{propA}-\ref {propB}
for the billiard. The Perron-Frobenius operator and its spectral
decomposition was studied in \cite{ClasGraph}. Here we use the
tools developed in that article. In particular, as explained in
\cite{ClasGraph}, the Pollicott-Ruelle spectrum $\{s_i\}$ for graphs is
obtained by computing the zeros of the secular equation
\begin{equation}
\det[I-Q(s)]=0,
\label{det}
\end{equation}
where, in the case that concerns us, the matrix $Q$ of the infinite  system
has elements $Q_{ab}=P_{ab}e^{sL_b}$, given by the probability of going
from the bond $b$ to the bond $a$ times the exponential function  involving
the length $L_b$ of bond $b$. Here we must take the bonds to be  directed,
thus $b$ and $\hat{b}$ represent the same bond, but with opposite
directions.  In our case, the graph has a linear structure and only
neighboring bonds are coupled. The matrix $Q$ is ordered in such a  way that
the column $a$ and row $b$ goes in order of increasing index,  alternating
directions, $\{\ldots,-1,\widehat{-1},0,\hat{0},1,\hat{1},\ldots\}$,
\begin{equation}
Q(s) = \left(\begin{array}{cccccc}
\ddots&&T_L e^{sL_1}&&&\\
&0&R_L e^{s L_1}&0&0&\\
&R_R e^{s L_1}&0&0&T_L e^{s L_2}&\\
&T_R e^{s L_1}&0&0&R_L e^{s L_2}&\\
&0&0&R_R e^{s L_2}&0&\\
&&&T_R e^{sL_2}&&\ddots
\end{array}\right)
\label{matrix}
\end{equation}
From Eqs.~(\ref{det})-(\ref{matrix}), and from the definition of $L_i $, we
infer the following property~:
\begin{property}[Analogous to property \ref{propA}]
\label{propA2}
If $s_0$ is a solution of Eq.~(\ref{det}) then so is $s_0\mu$.
\end{property}

To see this, note that changing $s$ by $s\mu$ in Eq.~(\ref {matrix})
amounts to shifting the elements of the matrix by two rows downwards and two
columns to the right, because it is equivalent to changing $L_i \to
L_{i+1}$. Since the matrix is infinite, this change leaves $Q$
unchanged. Thus,  if $s$ is a solution of the secular equation (\ref{det})
then so is $s\mu$. Thus, given a solution $s_0$, we can identify a
family of solutions  $\{s_i\}_{-\infty}^{\infty}$  where each $s_i=s_0\mu^i$. 
Given a different solution $s'_0$, there is another family 
$\{s'_i\}_{-\infty}^{\infty}$. 

For a given family of solutions,  the ratio between to roots of
Eq. (\ref{det}) is an entire power of $\mu$. Roots from different families
do not have this property. 

\begin{property}[Analogous to property \ref{propB}] \label{propB2}
If $\chi_0 = \{\ldots, a_{-1} a_{\widehat{-1}}, a_0, a_{\hat{0}}, a_1,
a_{\hat{1}},\ldots\}$ is the eigenvector associated to $s_0$,
{\em i.~e.} $Q(s_0)\chi_0=\chi_0$, then  the eigenvector associated to $s_1=s_0
\mu$ is $\chi_1 = \{\ldots, a_{0}, a_{\hat{0}}, a_1, a_{\hat{1}}, a_2,
a_{\hat{2}},\ldots \}$, whose elements are shifted to the left  compared to
$\chi_0$. That is, if $a_i$ is the $i$th element of $\chi_0$,
then it is also the $(i-2)$th element of $\chi_1$ or in  general
$\chi_{j+1}[b]=\chi_j[b+2]$, which implies,  $\chi_j[b]=\chi_0[b+2j]$.
\end{property}

With the same argument we have that the left eigenvector satisfies
$\tilde{\chi}_{j+1}[b]=\tilde{\chi}_j[b+2]$. 

Property \ref{propB2} has a simple
interpretation. First, note that due to the order we used for the matrix
$Q$ a shift by two elements on the vectors corresponds to a shift by one
cell on the  chain. Therefore, property \ref{propB2} says that, the mode
$\chi_{j+1}$ is equal to  the mode $\chi_j$, but shifted to the left  by one
cell. And property \ref{propA2} says that the mode $\chi_{j+1}$ has a
lifetime which is a factor  $1/\mu$ shorter, if $\mu>1$, (resp. longer if
$\mu<1$) than the lifetime of the mode $\chi_j$. Thus the spectrum of  the
evolution operator of the graph has the same properties as the  spectrum of
the evolution  operator of the billiard.

\subsection{From the Spectrum of self-similar Systems to the Drift }
\label{secDrift}

As was shown in numerical simulations of particles moving in the
billiard in Sec. \ref{sec_billnr}, a density of particles drifts towards
the  regions of growing cell size of the chain with a constant velocity,
while at the  same time it spreads ballistically. We will show shortly this
also holds for the graph.

A heuristic justification for the constant drift can be given thanks to the
self-similar structure of the system as follows. Anywhere in the system, the
probability is greater for particles to exit in the direction of  increasing
scales. Furthermore as the scales grow, the average time a particle  spends
in a cell increases accordingly. Hence we expect the drift to  remain
constant.

Now we offer a quantitative analysis of the drift, based upon properties 
\ref{propA} and \ref{propB}, or, equivalently, \ref{propA2} and
\ref{propB2}. 

The spectral decomposition of the Perron-Frobenius operator $\widehat{P}^t$
was obtained in  \cite{thomas,ClasGraph} for graphs and in
\cite{GaspB,Gas96} for billiards. For graphs, the time 
evolution of an observable $A[b,x_b]$ has the asymptotic expression  ($t \to
\infty$)
\begin{equation}
\langle A \rangle_t=\langle A|\widehat{P}^{t}\rho_{0}\rangle=\sum_{j}\langle
A|\psi_{j}\rangle e^{s_{j}t}\langle  \tilde{\psi_{j}}|\rho_{0}\rangle+ \ldots
\label{eq1}
\end{equation}
with
\begin{eqnarray}
\langle A|\psi_{j}\rangle &=& \sum_{b'=- \infty}^{\infty} \chi_{j}[b']
\frac{1}{l_{b'}} \int_{0}^{l_{b'}}e^{- s_{j}x/v}A[b',x]dx
\label{eq2},\\
\langle\tilde{\psi_{j}}|\rho_{0}\rangle &=& \frac{1}{N_j}
\sum_{b'=- \infty}^{\infty} \tilde{\chi}_{j}[b'] \int_{0}^{l_{b'}} e^ {s_{j}x/
  v}\rho_{0}[b',x]dx,
\label{eq3}
\end{eqnarray}
and
\begin{equation}
N_j=\sum_{b}l_{b}\tilde{\chi}_{j}[b]^{*}\chi_{j}[b].
\label{eq4}
\end{equation}
In Eq.~(\ref{eq1}) the dots stand for terms whose coefficients carry
subexponential time dependence which may arise from degeneracies of the
spectrum and can be neglected in the long time limit.
Similar expressions can be obtained for billiards with the same
conclusion \cite{GaspB,Gas96}, however for the sake of simplicity we will
state the results for graphs only.

We assume the initial density to be localized on one bond,
namely $\rho_{t=0}[b,x_b]= \delta_{b,b_{0}} \delta(x_b-x_{0})$, {\em
  i.~e.} all the particles 
start from $x_0$ on the bond $b_0$.
To simplify as much as possible the calculation and notation we will
consider $x_{0}=0$ 
and the bond $b_0$ as the reference bond $b_0=0$.
We are interested in the density as a function of the position
and $t$, {\em i. ~e.} $\rho_t[b,x_b]$, so we define the observable $A$ by
$A[b',x_{b'}] = \delta_{b,b'} \delta(x_{b}-x_{b'})$, in such a way  that
$\langle A 
\rangle_t=\rho_t[b,x_{b}]$, as we see from Eq.~(\ref{eq1}).  For
simplicity, we also take $x_b=0$, 
{\em i. e.} we measure the density at the beginning of the bond $b$.

Using this in Eq.~(\ref{eq1}), Eq.~(\ref{eq3}) and Eq.~(\ref{eq4}) we get
\begin{equation}
\langle
A\rangle_t=\sum_{j}\chi_{j}[b] \frac{e^{s_{j}t}}{N_j l_{b}}
\tilde{\chi}_{j}[0].
\label{sum-gen}
\end{equation}
Now, since the spectrum is divided in families according to whether  or  not
$s_j/s_k$ is an entire power of $\mu$, let us consider the  contribution of
only one of these families to the sum in Eq.~(\ref{sum-gen}).

From property \ref{propB2}, if $\chi_{j}[b]$ is the eigenstate  associated to $s_j$
then, the eigenstate associated to  $s_{j}/\mu$ is
$\chi_{j-1}[b]=\chi_{j}[b-2] $ and corresponds to the state  $\chi_{j}$
but shifted one cell to the right. It is then easy to see that
$N_j=\mu^{-2j}N_0$.  

Moreover, from Property \ref{propA2}, we have that if $s_0$ is a decay rate
(Pollicott-Ruelle resonance), then there is a family of Pollicott-Ruelle
resonances associated to it given by $s_j=s_0\mu^j$.

Now using the expressions of the lengths $L_{b}=L_0\mu^{b}$, we have that
the contribution to Eq.~(\ref{sum-gen}) due to this familiy is
\begin{equation}
\frac{1}{N_0} \sum_{j=-\infty}^{\infty} \chi_{j}[b]
\frac{\mu^{2j}}{\mu^b}e^{s_0\mu^jt}\tilde{\chi}_{j}[0]
\label{eq10}
\end{equation}

Let us split this sum in two parts, first the terms with
$j=-\infty,\ldots,0$ and second $j=1,\ldots,\infty$. We consider for now
the first term only, {\em i.~e.} $j=0,-1,\dots,-\infty$.

\begin{enumerate}
\item For $j=0,-1,\dots,-\infty$, $s_{0}$ is the largest decay
  rate. Therefore the component with the corresponding rate, $\chi_ {0}$, is
  the first to decay. After it has decayed, the  part of the density
  represented by this part of the sum --which we refer to as the  density
  for short--  moves to the right, because, as a  consequence of  property
  \ref{propB2},  all the other modes are shifted to the right of
  $\chi_{0}$.
\item The support of the density is shifted to the right by a  distance $L$,
  corresponding to the length of the bond where the
  mode $\chi_0$ is centered. Let us call it $b_{c}$. 
\item Then, the component of the mode $\chi_{-1}$ is the second to   decay
  with a rate $s_0/\mu$ and the support of the density moves another  bond
  to the right  because, by property \ref{propB2}, the bond where  the mode
  $\chi_{-1}$ is centered is the one at the right of $b_{c}$, {\em i. e.}  it is
  the bond $b_{c+1}$. 
  This bond $b_{c+1}$ is larger by a factor $\mu$ than $b_{c}$.
\end{enumerate}

Thus if we take the distance the packet moves divided by the
characteristic time (given by the inverse of the decay-rate), we have  that,
during the decay time of the first mode, the speed was $Ls_0$ and,  during
the decay time of the second mode to decay, the speed was
$(L\mu)s_0/\mu=Ls_0$, equal to the first. We can obtain the same  result for
the decay of the  third mode, $(L\mu^2)s_0/\mu^2=Ls_0$, and so on.

Now, it is
clear that the same conclusion follows from the other terms
$j=1,\ldots,\infty$ of the sum, thus we conclude the packet moves at a
constant  speed. 

This argument shows that a constant drift of particles has to be expected
in the self-similar systems as defined here.

If we were able to obtain the exact eigenvectors we could compute
numerically this sum and compare with simulations. Instead we propose a
heuristic argument and assume the eigenvectors are localized
around some finite region of the chain. A simple possibility is to assume a
Gaussian shape for the left and right eigenvectors, satisfying property
\ref{propB2}, namely $\chi_{j}[b]=\tilde{\chi}_{j}[b]=\exp[\frac{-(b-(b_{0}
  -2j))^{2}}{\sigma^{2}}]$.
The argument that the eigenvectors are localized can be justified with a
perturbative calculation provided the transmission probability $p\ll 1$,
similar to \cite{thomas}. 

For this heuristic calculation, we consider values of $\mu$ close to   one,
so we write $\mu=1+\epsilon$ and we can compute the sum in Eq.~(\ref {eq10})
for  small $\epsilon$. We also do the change of variable (we take  $L_0=1$)
$y=\frac{\mu^{b}-1}{\mu-1}\Rightarrow b\approx y+\epsilon y^{2}/2$
which measures the distance from the origin to the bond $b$. Then we   obtain
\begin{equation}
\langle A\rangle\sim
\exp[-\frac{y^{2}}{2\sigma^{2}}- \epsilon\frac{y^{3}}{2\sigma^{2}} +
s_{0}t(1-\epsilon y/4)-\frac{3y\epsilon}{2}]
\label{eq12}
\end{equation}
Now, we calculate the position of the maximum of Eq.~(\ref{eq12}) as a
function of $t$ and obtain
\begin{equation}
y_m\approx-\epsilon\frac{\sigma^{2}}{2}(s_{0}t/2+3)
\label{Gvel}
\end{equation}
We have a maximum that is linear in $\mu-1$, moves in the appropriate
direction and is zero if $\mu=1$, in agreement with simulations to be
presented next. Thus we have
shown that the two properties of the spectrum are related to the  motion  of
the packet.
It will be desirable to have a similar understanding of the  spreading  of
the density in terms of the spectrum of the Liouvillian operator.

\subsection{Transport Properties of the self-similar Graph: Numerical Results}
\label{num-graph}

\begin{figure}[h]
\centering
\includegraphics[width=.4\textwidth]{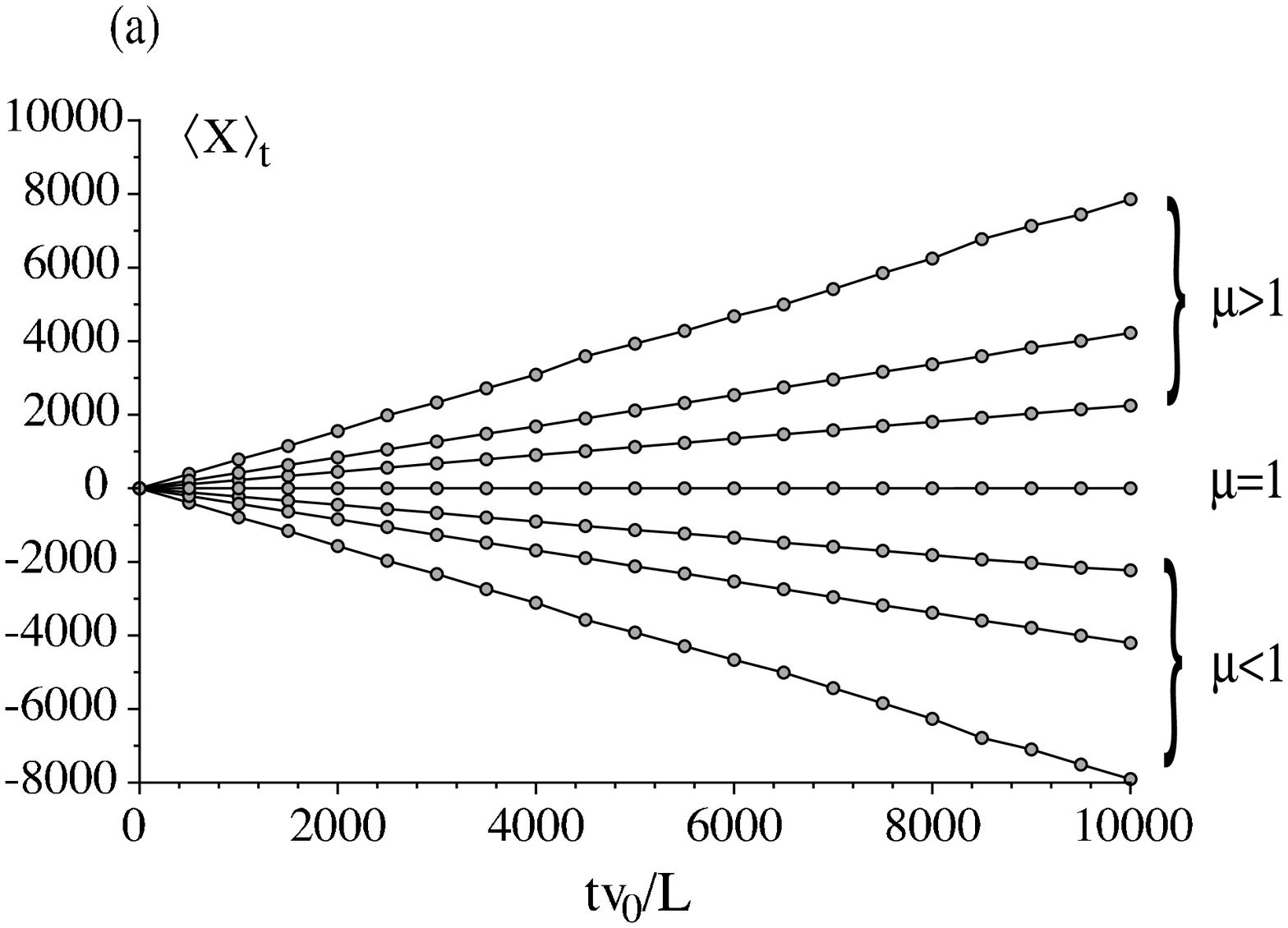}
\includegraphics[width=.4\textwidth]{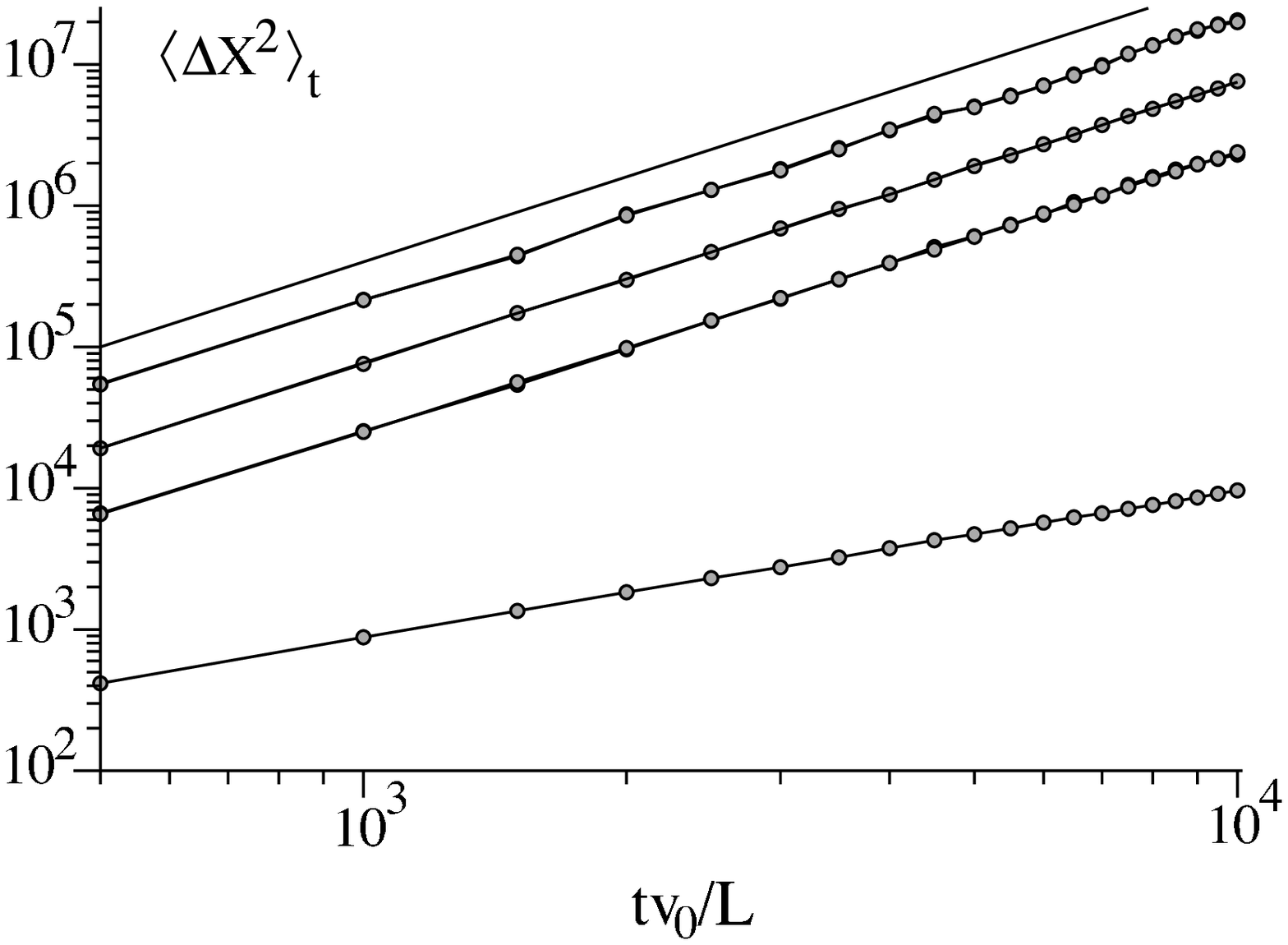}
\caption{(a) Average horizontal position $\langle X\rangle_t$  in the graph
  vs. time, for different values of $\mu$. We see that for each value of
  $\mu$ there is a well defined slope (velocity) and as $\mu$ increases so does the slope.
  (b) Mean square displacement $\langle \Delta X^2\rangle_t $,  for
  the same values of $\mu$. We observe a clear $\sim t^2$ (upper line) for
  all values of $\mu$ except for $\mu=1$  (lower curve) where the behavior
  is $\sim t$. The same values of $\mu$ as in Fig.~\ref{Fig-vel-billiard}
  were used.} 
\label{Figxvst}
\end{figure}

\begin{figure}[h]
\centering
\includegraphics[width=.4\textwidth]{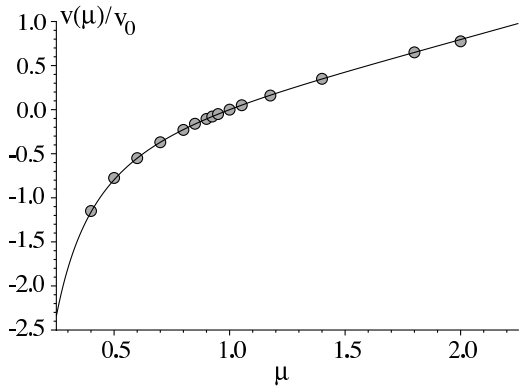}
\caption{Plot of the velocity $v$ as a function of $\mu$. The line  is  the
  function $c(\mu-1/\mu)(\sqrt{\mu}+1/\sqrt{\mu})$, with $c$ a   constant we
  use to fit the data. The reason to  plot
  this function will become clear in Sec.~\ref{SecMeq}}
\label{Fig5}
\end{figure}

The evolution of a density of particles in this system behaves in a way
similar to the billiard chain.
As with the billiard, a well-defined constant drift is observed, as
shown in Fig.~\ref{Figxvst} (a). The dependence of this drift in $\mu$ is
depicted in  Fig.~\ref{Fig5}. 
Here we also observe that the mean square displacement $\Delta  X^2= \langle
X^2\rangle-\langle X\rangle^2$ as a function of time, Fig.~\ref{Figxvst} (b),
behave like $\Delta X^2\sim t^2$. This ballistic  spreading of the  density is
also found with the macroscopic description of the system as given in
Sec. \ref{SecMeq}. 

\section{Macroscopic description}
\label{SecMeq}

In this
section we present a
system which is a charicature of the self-similar systems considered so far
and helps understand the macroscopic evolution on systems  with
self-similar structure. It describes the dynamics based on
a Master equation approach \cite{Hanson}. For this model we are able to
obtain analytical expressions for the (constant) drift velocity
and the ballistic spreading of the density.

We consider a discrete sequence of states
$i\in\mathbb{Z}$ and, associated to them, we have a conditional probability
$P_i(t)$ which represents the probability of being at site $i$ at  time
$t$ given some initial condition to be specified later.
We will identify the sequence of  sites
with the sequence of cells 
of the billiard or graph. Since for the billiard and graph only
transitions to neighboring cells are allowed, we introduce the   transition
rates $W^{-}_i$ and $W^{+}_i$, respectively for transitions to the left and
to the right of the site $i$. The master equation rules the time evolution
of the $P_i (t)$ and for this process it has the form
\begin{equation}
\partial_{t}P_i(t)=P_{i+1}(t)W^-_{i+1} + P_{i-1}(t)W^+_{i-1}-
P_i(t)(W^+_i + W^-_i),
\label{mastereq}
\end{equation}
It is a gain-loose or balance equation.
To mimic the transitions that occur on the graph, we consider  $W^{-} _i$
(respectively $W^{+}_i$) as the ratio between the probability of  going  to
the left, {\em i. e.} $p/\sqrt{\mu}$ (respectively $p\sqrt{\mu}$ for the right)
over the characteristic time associated to bond $i$ $L_0\mu^i/ v_0$, {\em
  i.~e.}, 
\begin{eqnarray}
W^{-}_i&=&\frac{pv_0}{L_0}\mu^{-i-1/2},
\label{tm}\\
W^{+}_i&=&\frac{pv_0}{L_0} \mu^{1/2-i}.
\label{tp}
\end{eqnarray}
Thus Eq.~(\ref{mastereq}) becomes
\begin{equation}
\partial_{t} P_{i}(t) = \frac{pv_0}{L_0}\mu^{-i}
\left[\mu^{-3/2}P_{i+1}(t)  + \mu^{3/2} P_{i-1}(t)  -
(\mu^{1/2} + \mu^{-1/2})P_{i}(t)\right].
\label{18}
\end{equation}
In matrix notation, this reads
\begin{equation}
\partial_t P_i(t) =
\mathcal{L}_{ij}P_j(t) \ ,
\label{Meq-M}
\end{equation}
where
\begin{equation}
\mathcal{L}_{ij}
= \frac{pv_0}{L_0}
\mu^{-j}[\mu^{-1/2}\delta_{j,i+1} + \mu^{1/2}\delta_{j,i-1}
-(\mu^{1/2}+\mu^{-1/2})\delta_{ij}].
\label{Lop}
\end{equation}
By exponentiation, since $P_j(t=0) = \delta_{j0}$, 
\begin{equation}
P_i(t) = \exp[\mathcal{L}t]_{ij}\delta_{j0}
\end{equation}
Note the symmetry of $\mathcal{L}$, $\mathcal{L}_{ij}(\mu) =
\mathcal{L}_{-ji}(1/\mu)$, implies $P_i(t)\to P_{-i}(t)$ upon
inverting $\mu$.

Note that until here, the model does not contain any metric
information. Only part of the 
self-similar structure is introduced through the transition rates $W^ {\pm}_i$.
The geometrical part of the self-similar structure is implemented
by associating a quantity $X_i$ to each site $i$ which is a measure  of the
distance from the origin to the site $i$.
We define it following the geometry of the graph.
If the site $i=0$ corresponds to the bond of length $L_0$,
and the site $i$ to the bond of length $L_i=L_0\mu^i$,
$X_i$ measures the distance from the middle of the $i$th bond to the
middle of the $0$th bond ({\em i.~e.} the origin), {\em
  viz.}
\begin{equation}
X_i(\mu)= \frac{L_0(1-\mu^i)}{2}\frac{1+\mu}{1-\mu}.
\label{Xi}
\end{equation}
In the expressions on the RHS, we observe the symmetry
$X_i\to-X_{-i}$ under the change $\mu\to1/\mu$. In other words
$X_i(1/\mu)=-X_{-i}(\mu)$.

We define the average position at time $t$ by
\begin{equation}
\langle X \rangle_t =  \sum_i P_i(t) X_i,
\end{equation}

The velocity is defined by
\begin{equation}
V(t) = \partial_t\langle X\rangle_t = \sum_{ij} \mathcal{L}_{ij} P_j (t)  X_i.
\end{equation}
According to the symmetries above, we should have $V\to-V$ upon   inverting
$\mu$, which we can check by direct calculation. We have
\begin{eqnarray}
V(t) &=& \frac{pv_0}{L_0}
\sum_j \mu^{-j} P_j [\mu^{-1/2} X_{j-1} + \mu^{1/2} X_{j+1}
- (\mu^{-1/2} + \mu^{1/2}) X_{j}],\nonumber\\
&=& \frac{pv_0}{L_0}
\sum_j \mu^{-j} P_j [\mu^{-1/2} (X_{j-1}-X_j) + \mu^{1/2}  (X_{j+1}- X_j)]\ .
\nonumber
\label{vel-2}
\end{eqnarray}
Using Eq.~(\ref{Xi}), we obtain the expression
\begin{equation}
V(t)=\frac{pv_0}{2}(\mu-\mu^{-1})(\mu^{-1/2}+\mu^{1/2}).
\label{vel}
\end{equation}
We observe that the speed has a constant value which vanishes for $\mu=1$
as expected. Moreover it is clearly anti-symmetric upon inverting
$\mu$.

Now we evaluate the time derivative of the mean square displacement,
\begin{equation}
\partial_t \langle \Delta X^2\rangle\equiv\partial_t[\langle  X^2 \rangle_t -
\langle X\rangle_t^2] =
\partial_t \langle X^2 \rangle_t -
2 \langle X \rangle_t \partial_t \langle X\rangle_t.
\label{tdermsq}
\end{equation}
with $\langle X^2\rangle_t=\sum_i P_i(t)
X_i^2$. Eq.~(\ref{tdermsq}) can be written as
\begin{equation}
\partial_t \langle \Delta X^2\rangle = \sum_{ij} \mathcal{L}_{ij}  P_j (t)X_i
[X_i-2\langle X\rangle_t].
\label{mean2-1}
\end{equation}
replacing $\mathcal{L}$ given in Eq.~(\ref{Lop}) in Eq.~(\ref {mean2-1})  we get
\begin{eqnarray}
\partial_t \langle \Delta X^2\rangle &=&
\frac{pv_0}{L_0} \sum_{ij}  \mu^ {-j} P_j(t)[\mu^{-1/2}
(X_{j-1}-X_j)(X_j+X_{j-1}-2\langle  X_t\rangle) + 
\mu^{1/2} (X_{j+1}-X_j)(X_{j+1}+X_{j}-2\langle  X_t\rangle)], 
\nonumber\\
&=&\frac{pv_0}{L_0}\left(\mu\sqrt{\mu}+\frac{1}{\mu\sqrt{\mu}}\right)
\left[\left(\frac{L_0}{2}\left(\sqrt{\mu}+\frac{1}{\sqrt{\mu}}\right)
\right)^2+\frac{v_0t}{2\sqrt{\mu}}\left(\mu-\frac{1}{\mu}\right)^2 \right],
\label{Dx2}
\end{eqnarray}
where we used Eq.~(\ref{Xi}) and the fact
that $\langle X_t\rangle=Vt$.

In the long time limit, {\em i.~e.} $t\gg t_{c}=\frac{L_0}{2v_0}\frac{\mu\sqrt
  {\mu}}{\mu-1}$, 
we can neglect the constant term and therefore $ \Delta X^2 \equiv  \langle
X^2 \rangle_t - \langle X\rangle_t^2\sim t^2$. In other words, the
spreading of the density is 
ballistic, similar to what we observed numerically in the billiard and
graph. In the 
opposite case $t\ll t_{c}$ 
the constant term dominates in $\partial_t \langle \Delta X^2\rangle$  and
the spreading is diffusive. 
The time $t_{c}=\frac{L_0}{2v_0}\frac{\mu\sqrt{\mu}}{\mu-1}$, marks the
crossover from diffusive to ballistic behavior of the mean-square
displacement.

Let us compare the speed, Eq.~(\ref{vel}), obtained in this macroscopic
model with 
the numerical results of sections \ref{sec_billnr} and \ref{num-graph}.
In the case of the billiard (see figure \ref{Fig-vel-mu-bill}), the agreement
is poor, because the Master equation is not a good model for the
billiard, except perhaps for $\mu\approx 1$. In fact, a particle in a
given cell 
of the billiard escapes  easily 
in a  few collisions and thus we do not expect that the master equation,
where the cell has a uniform probability, will
be a  good quantitative model. On the other hand what is observed is  that
for values of $\mu \approx 1$ the speed is proportional to
$\mu-1$.
In the case of the graph, see figure \ref{Fig5},
 Eq.~(\ref{vel}) fits very  well the data because  the
probability of being in a bond is well approximated  by a 
uniform  distribution in the limit $p\ll 1$ where many 
collisions occurs in average with the scatterers before the particle 
goes to another bond.
Thus we expect the master equation to be a good
model for this system.

Finally we show that for this macroscopic description, we also have
properties similar to \ref{propA} and \ref{propB} (or \ref{propA2} and \ref{propB2}), and
therefore a correspondence between the constant drift and the
spectral properties of $\mathcal{L}$.

The decay rates can be computed assuming a solution of the  form
$P_i(t)=e^{st}Q_i$ \footnote{The complete solution takes the form $P_i (t)
  \sim \sum_j e^{s_j t}Q^j_i+\cdots$,
where extra terms represented by the dots are expected from Jordan
Blocks.}. 
Upon substitution of this expression into Eq.~(\ref{Meq-M}), we find that
the decay rates $s$ must be solutions of
\begin{equation}
\det[s\delta_{ij}- \mathcal{L}_{ij}]=0 \ .
\label{m-det}
\end{equation}

Now we proceed as with the billiard and the graph.
We assume a particular solution $s$ of Eq.~(\ref{m-det}) was  obtained
and analyze 
$\det[s\mu\delta_{ij}- \mathcal{L}_{ij}]$. Note that due to the definition
of $\mathcal{L}$ we have $\mathcal{L}_{i,j}=\mu\mathcal{L}_{i+1,j+1}$.
Therefore,
\begin{equation}
\det[s\mu\delta_{ij}- \mathcal{L}_{ij}]=\det[\mu(s\delta_{i+1,j+1}-
\mathcal{L}_{i+1,j+1})]\ . 
\end{equation} 
Because $\mu\neq 0$, 
we conclude that if $s$ is a  solution of
Eq.~(\ref{m-det}) 
then so is $s\mu$.

Now, the associated eigenvector $Q^s_i$ is obtained from
$\sum_j[s\delta_{ij}- \mathcal{L}_{ij}]Q^s_j=0$ and
$Q^{s\mu}_i$ is obtained from $\sum_j[s\mu\delta_{ij}-
\mathcal{L}_{ij}]Q^{s\mu}_j=0$. Again, using the definition
of $\mathcal{L}$, it is easy to show that $Q^{s\mu}_i=Q^s_{i+1}$. This
shows we have again the equivalent of properties \ref{propA} and
\ref{propB} (or \ref{propA2} and \ref{propB2}).

\section{conclusions}
\label{conclu}

In this article we have studied the statistical properties of three
different classes of self-similar systems and found that all three share
very similar macroscopic properties;
we observe a drift of particle densities towards the direction of growing
scales at constant velocity, as well as a ballistic spreading of the
density. 

A justification for the presence of a drift was provided, for both the
billiard chains and graphs, in terms of two essential properties of the
spectrum of the evolution operator, the first stating that for
every decay-rate $s$, there is also a decay rate $s\mu$, and, correspondingly,
the second, that the eigenstate associated to $s\mu$ is shifted one cell to
the left of the chain  with respect to the eigenstate associated to $s$.   

As of the Billiard, we should note that these properties
hold independently of the exact  shape of the unit cell; they are merely a
consequence of the self-similar structure. The 
only relevant restriction regarding the shape of the unit cell is that the
dynamics in that billiard cell must be strongly chaotic in order to 
have the kind of spectral decomposition that we assumed. 

For the self-similar graph, these properties are also rather general. Here we
considered a very simple self-similar graph, to avoid 
complicated expressions, but we can take a unit cell composed of  any  number
of bonds, say $N_B$, and by ordering the matrix $Q$ in such a way  that all
the bonds of the same  cell are consecutive we will obtain property  
\ref{propA2} and property \ref{propB2}, which will look like
$\chi_{j+1}[b]=\chi_j[b+2N_B]$. 

Let us discuss an interpretation of these two properties in term of
periodic orbits. Consider the billiard. Decay rates and
eigenstates are determined by an ensemble of periodic orbits that form a
repellor \cite{Gas96}. Every periodic orbit can be shifted by one cell
to the right or to the  left and generate a new periodic orbit with the
only difference that the period is a factor of $\mu$ respectively shorter or
longer. If the periodic orbits determine the support of the eigenstates,
then it is clear that eigenstates of the same "shape" but shifted along
the billiard exist and that the associated decay rates will also differ by 
a factor of $\mu$. The  change in period (time scale) corresponds to property
\ref{propA} and the fact that the geometry of the orbits is unchanged to
property \ref{propB}.   

We have also provided a macroscopic description of self-similar billiards and graphs
in terms of a  Master Equation, for which we obtained an analytical
expression of the velocity and ballistic spreading. Properties equivalent to
\ref{propA} and \ref{propB} are also shared by this system.

We note that in going from the billiard to the graph and then to the
Master Equation, the stochasticity increases.
Billiards are deterministic chaotic systems; in graphs, there is a
stochastic element which acts only when the particles
reach the vertices, while the motion within the bond is deterministic; The
dynamics described by the Master Equation is completely stochastic with
transitions possibly occuring at any time. 

We should also note that the existence of a drift is rather
intuitive and  simple to explain qualitatively in terms of the
probabilities of  crossing to the left or right and in terms of the time
spent in the bigger or smaller cells. However, our aim was to provide an
example where a macroscopic property like the drift of particles can be
related to the microscopic  properties of the system, {\em i.~e.} the exact
Liouvillian evolution operator. 

Several perspectives are open for future research, of which we now mention
two. First,  it will be interesting to characterize the spreading in
terms of properties of the spectrum and, second, regarding the billiard, we
would like to obtain further understanding of the function $v(\mu)$. Our
main observation of this work is that there is a well-defined 
speed of propagation for each value of $\mu$. However, a determination of
the speed $v$ vs. $\mu$ is missing thus far.
Numerical results (not all presented here) show that, as we change the
parameters of the billiard, the speed displays a rich behavior as a
function of $\mu$. For instance, we observed functions $v(\mu)$ that are
monotonous with $\mu$ in some parameter regions, while in other regions,
such as corresponding to the parameter values used in
Fig.~\ref{Fig-vel-mu-bill} we observed a non-monotonous function.
Near $\mu=1$ the behavior of $v(\mu)$ is determined by the symmetry 
$v(\mu)=-v(1/\mu)$. In fact, it follows from that symmetry, that for small
values of  
$\mu-1$, we can expand $v(\mu)=v'(1)(\mu-1-(\mu-1)^2/2)+\ldots$  

We expect to continue our work in these directions. 
In particular, we believe a better understanding of the
dynamical system in the unit cell with boundary conditions given by
Table~\ref{table1} is needed, specifically regarding the
properties of the invariant measure. Because of the open boundary
condition, the invariant measure is not the Liouville measure. The volumes
are not conserved by the matching conditions. One therefore expects fractal
properties of the invariant measure. This will be the subject of further
publications. 

\appendix

\section{Finite Horizon self-similar Billiard Channel \label{Appgeomcell}}

In this appendix, we extend the finite horizon condition for the usual
periodic Lorentz gas to the self-similar case, so as to ensure
hyperbolicity of the dynamics.

In the absence of self-similar structure ($\mu=1$), the
Lorentz channel \cite{GaspB} has the symmetry of a two-dimensional periodic
Lorentz gas on a hexagonal lattice. This system is diffusive provided it
verifies the so-called finite horizon conditions, namely 
\begin{equation}
2 R < D < 4R/\sqrt{3}, 
\label{lgfh}
\end{equation}
where $R$ denotes the common radius of the scatterering disks
and $D$ the distance between neighboring disks. The lower bound ensures
that the disks do not overlap and the upper one that there are no free
flying trajectories. 

For a mixed Lorentz gas, with
alternating rows of scatterers of radii $r$ and $R$, the hexagonal
structure of the lattice is preserved, but the corresponding channel now
consists of three rows of scatterers~: the upper and lower rows with half
disks of radii $R$ and the middle row with disks of radii $r$. The
corresponding unit cell has a rectangular shape with four quarter disks of
radii $R$ at the corners and one disk of radius $r$ at the center 
(where the diagonals intersect), with coordinates
\begin{eqnarray}
4\ \mathrm{disks}\ \mathrm{of}\ \mathrm{radii}\ R&:&
\begin{array}{l}
\big(0,\pm\sqrt{3}D/2\big),\\
\big(D,\pm\sqrt{3}D/2\big),
\end{array}
\\ 
1\ \mathrm{disk}\ \mathrm{of}\ \mathrm{radius}\ r&:&(D/2,0).
\label{lorgaspos}
\end{eqnarray}
For this system, the finite horizon conditions become
\begin{eqnarray}
D &>& 2r,\label{mlgfh1}\\
D &>& 2R,\label{mlgfh2}\\
D &<& 2(r + R)/\sqrt{3}.\label{mlgfh3}
\end{eqnarray}

Now introducing the parameter $\mu$, we deform the cell of the
Lorentz channel from a rectangle to a trapezoid with vertical sides scaled
by $1/\sqrt{\mu}$ and $\sqrt{\mu}$ respectively, as was shown in
Fig. \ref{Fig1}. The positions of the disks become
\begin{eqnarray}
4\ \mathrm{disks}\ \mathrm{of}\ \mathrm{radii}\ R&:&
\begin{array}{l}
\big(0,\pm\sqrt{3}D/(2\sqrt{\mu})\big),\\
\big(D,\pm\sqrt{3}D\sqrt{\mu}/2\big),
\end{array}
\\ 
1\ \mathrm{disk}\ \mathrm{of}\ \mathrm{radius}\ r&:&(D/(1+\mu),0),
\end{eqnarray}
where the disk on the horizontal axis lies at the intersection of the
diagonals. 

Let us assume $\mu>1$ in the sequel.

\subsection{Non-Overlapping Disks}

The non-overlapping conditions Eqs.~(\ref{mlgfh1}-\ref{mlgfh2}) transpose
to three new conditions~: 
\begin{enumerate}
\item The sum of the radii of the external disks must be less than the
  length of the lower/upper side of the cell,
\begin{eqnarray}
\frac{R}{\sqrt{\mu}} + \sqrt{\mu}R &<&
\sqrt{D^2 + \left[\frac{\sqrt{3\mu}D}{2} -
    \frac{\sqrt{3}D}{2\sqrt{\mu}}\right]^2},\nonumber\\
&=& D\sqrt{\frac{3}{4}(\mu + \mu^{-1}) - \frac{1}{2}}.
\label{cond1}
\end{eqnarray}

\item The shortest diagonal must be greater than the sum of the radii
  $R/\sqrt{\mu}$ and $r$,
\begin{equation}
\frac{R}{\sqrt{\mu}} + r <
\sqrt{\left(\frac{\sqrt{3}D}{2\sqrt{\mu}}\right)^2 +
  \left(\frac{D}{\mu+1}\right)^2}.
\label{cond2}
\end{equation}

\item The radius of the center disk must be
    less than the distance from the intersection of the diagonal to the
    cell's boundary. That is,
\begin{equation}
r < \frac{D}{1+\mu}.
\label{cond3}
\end{equation}
\end{enumerate}

\begin{figure}[htb]
\centering
\includegraphics[width=.4\textwidth]{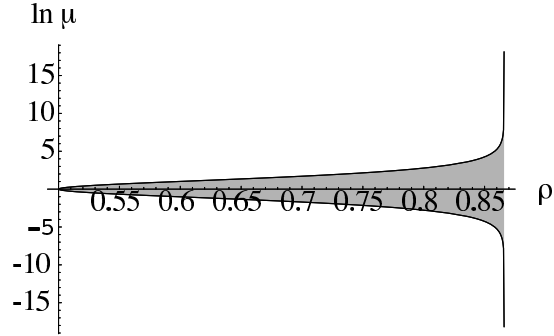}
\caption{From Eqs. (\ref{mu1}, \ref{cond13}). The gray region is forbidden
  when $1/2<\rho<\sqrt{3}/2$.}
\label{Fig1append}
\end{figure}

Let $\rho \equiv R/D$ and $\lambda \equiv r/D$ be two dimensionless
parameters. The first of the three conditions above, Eq.~(\ref{cond1}), can
be transformed into a second degree polynomial in $\rho$~:
\begin{equation}
(3/4-\rho^2)\mu^2 - (1/2 + 2\rho^2)\mu + (3/4 - \rho^2) > 0.
\label{cond12}
\end{equation}
We must assume
\begin{equation}
\rho < \frac{\sqrt{3}}{2}.
\end{equation}
The two roots of Eq.~(\ref{cond12}) are
\begin{equation}
\mu_\pm = 
\frac{1 + 4\rho^2 \pm 2\sqrt{8\rho^2 - 2}}{3-4\rho^2},
\label{mu1}
\end{equation}
which are real if $\rho \ge 1/2$. In this case, the condition (\ref{cond1})
is only satisfied provided either of 
\begin{equation}
\begin{array}{c}
\mu > \mu_+,
\mu <\mu_-,
\end{array}
\label{cond13}
\end{equation}
hold. Note that $\mu_+ = 1/\mu_-$. These conditions are depicted in
Fig. \ref{Fig1append}. 

Otherwise, if $\rho < 1/2$, Eq. (\ref{cond12}) has no real roots and 
Eq. (\ref{cond1}) is always true. 

Consider the second condition, Eq. (\ref{cond2}), which
takes the form
\begin{equation}
\left[\frac{3}{4}-(\rho + \lambda \sqrt{\mu})^2\right](\mu + 1)^2 + \mu >
0.
\label{cond21}
\end{equation}
This is a sixth order polynomial in $\sqrt{\mu}$, with potentially as many
roots, but is easy to solve numerically.

With these notations, we rewrite Eq. (\ref{cond3}) as
\begin{equation}
\mu < \frac{1}{\lambda}- 1.
\label{cond31}
\end{equation}

\subsection{Finite Horizon}

\begin{figure}[htb]
\centering
\includegraphics[width=.4\textwidth]{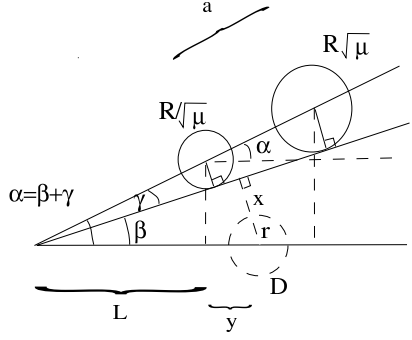}
\caption{Geometry and definition of $\alpha,\beta,\gamma$ and $x$} 
\label{fig.fhh}
\end{figure}

The condition that there are no free-flying trajectories, as seen in
Fig.~\ref{fig.fhh}, is that the line tangeant to the row of upper disks
intersect the disk at the center.

Consider the reference cell and let $x$ denote the distance from the center
of the central disk to the trajectory tangent to the upper disks. 
Since these lines are perpendicular, we can write 
\begin{equation}
x=(L+y)\sin\beta,
\label{xepr}
\end{equation}
where $y=D/(\mu+1)$ is the distance from the boundary of the cell to the
center of the disk and $L=D\sum_{-\infty}^{-1}\mu^i=D/(\mu-1)$.
In order to compute the angle $\beta$, we write $\beta=\alpha-\gamma$,
where $\alpha$ and $\gamma$ are given as follows.
 
First we have that the difference between the two verticals from the base
line to the upper right and upper left disks is
$\sqrt{3}/(2D)(\sqrt{\mu}-1/\sqrt{\mu})$ and $D$ is 
the horizontal distance between their centers. Therefore
\begin{equation}
\tan\alpha=\frac{\sqrt{3}}{2}\left(\sqrt{\mu}-\frac{1}{\sqrt{\mu}}\right)
\end{equation}
We note that the distance $a$ along the wall is given by  
\begin{eqnarray}
a&=&D/\cos\alpha, \nonumber\\
&=& \frac{D}{2}\sqrt{3(\mu+\mu^{-1}) - 2}.
\end{eqnarray}
We can then compute $\gamma$ from the equality
\begin{eqnarray}
\sin\gamma &=& \frac{R/\sqrt{\mu}}{a\sum_{-\infty}^{-1}\mu^i},
\nonumber\\
&=& \frac{R(\mu-1)}{a\sqrt{\mu}},\nonumber\\
&=& \frac{2R}{D}\frac{\mu-1}{\sqrt{3 \mu^2 -2 \mu + 3}}.
\end{eqnarray}

Now, since $\beta=\alpha-\gamma$ we have
\begin{eqnarray}
\sin\beta &=& \sin\alpha\cos\gamma-\cos\alpha\sin\gamma,
\nonumber\\
&=& \frac{2\rho\sqrt{\mu}(\mu-1)}{3-2\mu+3\mu^2}\\
&&\times
\left[\sqrt{\frac{3}{\mu}\left[\frac{3-2\mu+3\mu^2}{4\rho^2} -
      (\mu-1)^2\right]}-2\right]. \nonumber
\end{eqnarray}

The finite horizon condition is
\begin{equation}
\lambda = \frac{r}{D}>\frac{2\mu}{\mu^2-1}\sin\beta.
\label{cond4}
\end{equation}
In the limit $\mu\to1$ one retrieves Eq. (\ref{mlgfh3}).

\subsection{Choice of Parameter values}

We want to fix the values of $\rho = R/D$ and $\lambda = r/D$ and find the
range of values of $\mu$ so as to verify the constraints specified by
Eqs.~(\ref{cond1}, \ref{cond2}, \ref{cond3}, \ref{cond4}). 

\begin{figure}[htb]
\centering
\includegraphics[width=.4\textwidth]{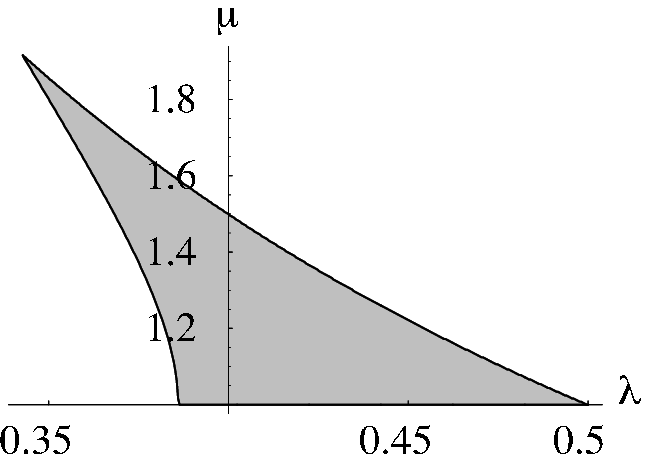}\quad
\includegraphics[width=.4\textwidth]{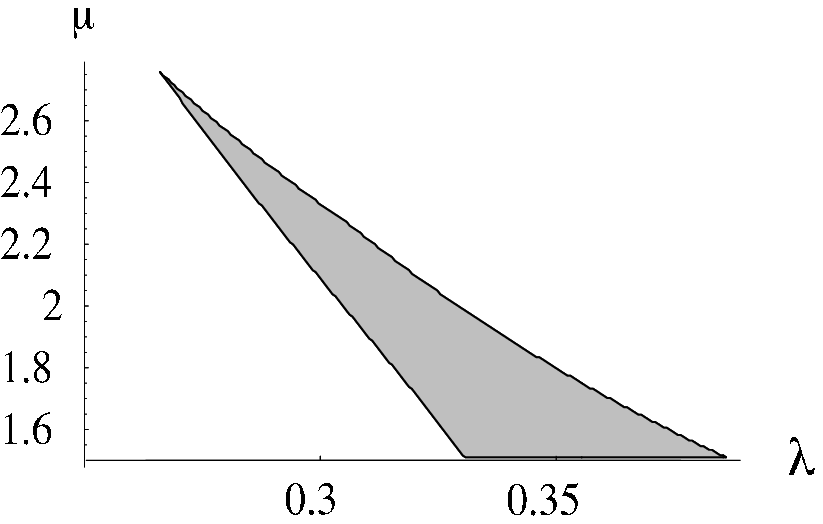}
\caption{Values of $\mu$ vs. $\lambda$, for fixed $\rho$ = 0.48 (a), 0.52
  (b), which are consistent with Eqs.~(\ref{cond1}), (\ref{cond2}),
  (\ref{cond3}) and (\ref{cond4}).} 
\label{fig.param1}
\end{figure}

\begin{figure}[htb]
\centering
\includegraphics[width=.4\textwidth]{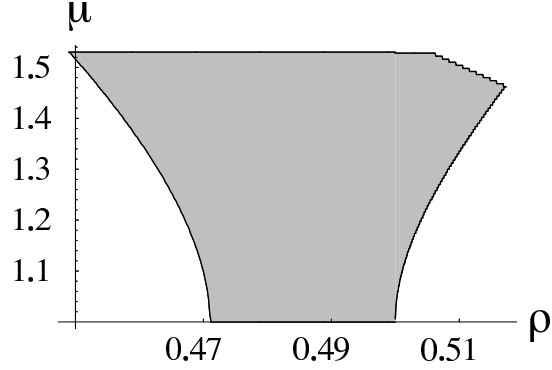}
\caption{Values of $\mu$ vs. $\rho$, for fixed $\lambda$ = 0.395 which are
  consistent with Eqs.~(\ref{cond1}), (\ref{cond2}), (\ref{cond3}) and (\ref{cond4}).} 
\label{fig.param2}
\end{figure}

The numerical results presented in Sec.~\ref{sec_billnr} use the fixed
parameters $\lambda=r/D=0.395$ and $\rho=R/D=0.480$. For those values, the
range of allowed values of $\mu$ is $0.653\le\mu\le1.532$. For the sake of
illustration, other possible parameter values are shown in
Figs.~\ref{fig.param1}-\ref{fig.param2} with either $\rho$ or $\lambda$
fixed.   

\section{Alternative proof of properties \ref{propA} and \ref{propB} for a  self-similar
  billiard \label{appendix}}

The dynamics in the billiard chain is Hamiltonian and therefore the
Liouville measure is invariant under the time evolution.
According to Birkhoff, the Liouville measure can be expressed as  $d \zeta
d\tau$, where $\zeta=(s,v_\mathrm{t})$ is a pair of variables with $s$
representing the arc-length of the billiard (not the unit cell), and
$v_\mathrm{t}$ 
the projection of the velocity to the tangential direction at the  point
$s$. Because the energy is conserved we can fix the speed $|v|=1$. With
this  choice $-1<v_\mathrm{t}<1$. Now, $\tau$ is a variable which measures a
distance along a line drawn from the point $s$ with the direction  given  by
$v_\mathrm{t}$. Obviously, $\tau$ can vary from zero to the intersection of
this 
line with the billiard. This distance will be denoted by  $T(\zeta)$,
{\em i.~e.} $0<\tau<T(\zeta)$. Thus we can consider that the average of a
function 
$f(\zeta,\tau)$ is  given by
\begin{equation}
\langle f \rangle=\frac{1}{\int ds \int dv_\mathrm{t} \int_0^{T(\zeta)}
  d\tau} \int 
ds \int dv_\mathrm{t} \int_0^{T(\zeta)} d\tau f(s,v_\mathrm{t},\tau).
\end{equation}
The variable $s$ is unbounded because the billiard is infinite. Due  to  the
construction of our billiard, we can split the integral over the arc-length
into the contributions over the differents cells, {\em i.~e.}
\begin{equation}
\int ds=\sum_I \int_0^{L_I}ds \ .
\end{equation}
Due to the symmetry $L_I=\mu^I L_0$, we can let $0<s<L_0$. The   restriction
of the variable $\zeta$ to this region will be called  $\xi$. We also
observe that $T(\zeta)=T(\xi,I)=L(\xi)\mu^I$. Thus the average  can be
written as
\begin{equation}
\langle f \rangle=\frac{1}{\sum_I \mu^I\int ds \int dv_\mathrm{t}
  \int_0^{T (\xi,I)} 
  d\tau}\sum_I \mu^I\int_0^{L_0} ds \int dv_\mathrm{t}  \int_0^{T(\xi,I)} d\tau
f(s,v_\mathrm{t},\tau,I),
\end{equation}
or, in a more compact form, using $\int d\xi T(\xi,I)=\mu^I\langle
L(\xi)\rangle$, 
\begin{equation}
\langle f \rangle=\lim_{J\to\infty}\frac{1}{\sum_{I=-J}^{J}
  \mu^{2I}\langle L(\xi)\rangle}\sum_{I=-J}^{J} \mu^I\int d\xi
\int_0^{T(\xi,I)} d\tau f(\xi,\tau,I).
\end{equation}
Now, the quantity that is of interest to us is
\begin{eqnarray}
\langle \tilde{b}_{s^*}b_s\rangle
&=&\lim_{J\to\infty}\frac{1}{\sum_{I=-J}^{J}
  \mu^{2I}\langle L(\xi)\rangle}\sum_{I=-J}^{J} \mu^I\int d\xi
\int_0^{T(\xi,I)} d\tau \tilde{b}_{s^*}(\xi,I)b_s[\xi,I],\nonumber\\
&=& \lim_{J\to\infty}\frac{1}{\sum_{I=-J}^{J}
  \mu^{2I}\langle L(\xi)\rangle}\sum_{I=-J}^{J} \mu^{2I}\int d\xi L (\xi)
\tilde{b}_{s^*}(\xi,I)b_s[\xi,I].
\label{eq-s}
\end{eqnarray}
We assume that $s$ is a resonance {\em i.~e.}, $\langle
\tilde{b}_{s^*}b_s\rangle=1$ \cite{Gas96} and we want to show that this implies
$\langle \tilde{b}_{\mu s^*}b_{\mu s}\rangle=1$, that is that $s \mu $  is
also a resonance.

To show this, we note that, from the formal expression of $b_s$, we have
$b_{\mu s}[\xi,I]=b_s[\xi,I+1]$, and therefore
\begin{equation}
\langle \tilde{b}_{\mu s^*}b_{\mu s} \rangle =
\lim_{J\to\infty} \frac{1}{\sum_{I=-J}^{J} \mu^{2I}\langle 
  L(\xi)\rangle}\sum_{I=-J}^{J} \mu^{2I}\int d\xi L(\xi)
\tilde{b}_s[\xi,I+1]b_s[\xi,I+1],
\end{equation}
which is equal to
\begin{equation}
\langle \tilde{b}_{\mu s^*}b_{\mu
  s}\rangle=\lim_{J\to\infty}\frac{1}{\sum_{I=-J+1}^{J+1} \mu^{2I} \langle
  L(\xi)\rangle}\sum_{I=-J+1}^{J+1} \mu^{2I}\int d\xi L(\xi)
\tilde{b}_s[\xi,I]b_s[\xi,I].
\label{eq-s-mu}
\end{equation}
If the limit in Eq.~(\ref{eq-s}) exists (as we asume), then the
limit in Eq.~(\ref{eq-s-mu}) also exists and they are equal. Therefore we
conclude 
that $\langle \tilde{b}_{\mu s^*}b_{\mu s}\rangle=1$ and so $s\mu$ is  also  a
resonance. This proves property (A).

Now, from the formal expression for the eigenstate associated to $s$,
\begin{equation}
b_s[\xi,I]  =\prod_{j=1}^{\infty}\exp[-sT(\phi^{-j}\xi, I -
\sum_{i=1}^{j}a(\phi^{- i}\xi))],
\label{eq15}
\end{equation}
we can express $b_{s\mu}[\xi,I]$, the eigenstate associated to $s\mu$ as
\begin{eqnarray}
b_{s\mu}[\xi,I] &=& \prod_{j=1}^{\infty} \exp[-s\mu  T(\phi^{-j}\xi, I -
\sum_{i=1}^{j}a(\phi^{-i}\xi))],\nonumber\\ 
&=& \prod_{j=1}^{\infty} \exp[-sT(\phi^{-j}\xi, I +1 - \sum_{i=1}^{j}
a(\phi^{-i}\xi))],\nonumber\\
&=&b_s[\xi,I+1],
\end{eqnarray}
where the second equality follows from the definition of $T(\xi,I)$ in
Eq.~(\ref{Time}) and the last from Eq.~(\ref{eq15}). This is to say that the
eigenstate associated to $s\mu$ is the same than the one associated  to $s$
but shifted one cell to the left, and proves property (B).

\begin{acknowledgments}
FB acknowledges financial support from Fondecyt under project  1030556. TG
is charg\'e de recherches with the Fonds National de la Recherche
Scientifique (Belgium).
\end{acknowledgments}

\end{document}